\documentclass[a4paper]{article}
\usepackage{geometry}
\usepackage{amsmath}
\usepackage{amssymb}
\usepackage{indentfirst}
\usepackage{mathdots}
\usepackage{cite}
\usepackage[colorlinks, linkcolor=black, citecolor=blue]{hyperref}
\usepackage{mathrsfs}
\usepackage{booktabs}
\usepackage{enumitem}
\usepackage{authblk}
\usepackage{amsfonts,ntheorem}

\makeatletter

\newcommand{\Rmnum}[1]{\expandafter\@slowromancap\romannumeral #1@}
\makeatother

\newtheorem{theorem}{Theorem}[section]
\newtheorem{lemma}{Lemma}[section]
\newtheorem{corollary}{Corollary}[section]
\newtheorem{proposition}{Proposition}[section]
\newtheorem{problem}{Problem}[section]

{ 
\theoremstyle{plain}
\theorembodyfont{\normalfont} 

\newtheorem{definition}{Definition}[section]
\newtheorem{example}{Example}[section]
\newtheorem{remark}{Remark}[section]

}

\newenvironment{proof}{\noindent{\textbf{\emph{Proof.}}}}

\allowdisplaybreaks[4]

\begin {document}
\title{\Large Quantum error-correcting codes from matrix-product codes related to quasi-orthogonal and quasi-unitary matrices}
\author[1,2]{{\small{Meng Cao}} \thanks{E-mail address: mengcaomath@126.com}}
\affil[1]{Beijing Institute of Mathematical Sciences and Applications, Tsinghua University, Beijing, 101408, China }
\affil[2]{Yau Mathematical Sciences Center, Tsinghua University, Beijing, 100084, China }

\date{}
\maketitle

\begin{abstract}
Matrix-product codes over finite fields are an important class of long linear codes by combining several commensurate shorter linear codes
with a defining matrix over finite fields.
The construction of matrix-product codes with certain self-orthogonality over finite fields is an effective way to obtain good $q$-ary
quantum codes of large length.
This article has two purposes: the first is
to summarize some results of this topic obtained by the author of this article and his cooperators in  \cite{Cao2020Constructioncaowang,Cao2020New,Cao2020Constructionof}; the second is to add some new results
on quasi-orthogonal matrices (resp. quasi-unitary matrices), Euclidean dual-containing (resp. Hermitian dual-containing)
matrix-product codes and $q$-ary quantum codes derived from these matrix-product codes.
\end{abstract}

\vspace{10pt}


\vspace{6pt}

\section{Introduction}
It is well known that quantum error-correcting codes (quantum codes, for short) are indispensable to quantum computation and quantum communication.
They were introduced to deal with the problems of decoherence and quantum noise in quantum information.
After the pioneering research in \cite{Calderbank1998Quantumerror,Shor1995Scheme,Steane1996Multiple-particle},
the theory of quantum codes has experienced a rapid development over the past two decades.
A great deal of research are focused on finding quantum codes with good parameters.
Usually, we use the notation $[[n,k,d]]_{q}$ to represent a $q$-ary quantum code with length $n$, dimension $q^{k}$ and minimum distance $d$.
It has the abilities to detect up to $d-1$ quantum errors and correct up to $\lfloor \frac{d-1}{2}\rfloor$ quantum errors.
Naturally, we know that when fixing both the length $n$ and the dimension $q^{k}$ (or $k$),
the larger value of $d$ means the better performance of error detection and error correction of the quantum code.

As we know, constructing quantum codes with good parameters is significant and difficult.
The CSS construction (see Theorem \ref{theorem3.1} for details),
introduced by Calderbank and Shor \cite{Calderbank1996Goodq} and Steane \cite{Steane1996Simpleq},
is a powerful method to construct $q$-ary quantum codes from classical codes with self-orthogonality over the finite field $\mathbb{F}_{q}$.
To be more specific, any Euclidean self-orthogonal or Euclidean dual-containing code over $\mathbb{F}_{q}$ will produce a $q$-ary quantum code.
For more information on CSS construction and its applications,
see \cite{LaGuardia2009,LaGuardia2014,Zhu2018Aclassof,Aly2007,Grassl2004,Grassl2015,Ezerman2013}.

In 2001, Ashikhmin and Knill \cite{Ashikhmin2001} presented another effective method, i.e., Hermitian construction
(see Lemma \ref{lemma4.1} for details),
for constructing $q$-ary quantum codes from classical codes over the finite field $\mathbb{F}_{q^{2}}$.
The Hermitian construction reveals that a Hermitian dual-containing $[n,k,d]$ code over $\mathbb{F}_{q^{2}}$
can yield a $q$-ary $[[n,2k-n,\geq d]]$ quantum code.
The Hermitian construction is currently one of the most frequently-used methods
for constructing good $q$-ary quantum codes from classical codes such as cyclic codes
(e.g., see \cite{Aly2007,Kai2012,Kai2014,Zhang2015Some,Chen2015}),
generalized Reed-Solomon (abbreviated to GRS) codes
(e.g., see \cite{Li2008,Jin2014A,Jin2017Quantum,Zhang2017})
and matrix-product codes
(e.g., see \cite{Cao2020Constructioncaowang,Cao2020Constructionof,Liu2018On,Liu2019Entanglement-assisted,Zhang2015qc,Jitman2017}).
In brief, both CSS construction and Hermitian construction establish the close relationship between classical codes and quantum codes,
which enable us to construct new quantum codes in a convenient way.

The matrix-product code
\begin{equation*}
C(A)=[C_{1},C_{2},\ldots,C_{k}]A
\end{equation*}
over finite fields is an interesting classical linear code with larger length by combining several commensurate linear codes
$C_{1},C_{2},\ldots,C_{k}$, i.e., constituent codes with the same length, with a defining matrix $A$ over finite fields.
This concept was proposed by Blackmore and Norton \cite{Blackmore2001} in 2001, the same year that the Hermitian construction
was proposed. Let $C_{i}$ be an $[n,t_{i},d_{i}]_{q}$ linear code for $i=1,2,\ldots,k$ and $A$ be an $k\times s$ matrix over
$\mathbb{F}_{q}$ with $k\leq s$. It is known from \cite{Blackmore2001} that the matrix-product code $C(A)$ has length $sn$ and dimension
$\sum_{i=1}^k t_i$ (see also \cite{O2002Note}).
Besides, the minimum distance of $C(A)$, denoted by $d[C(A)]$, is investigated in the following cases.
\begin{itemize}

\item In \cite{Blackmore2001}, Blackmore and Norton proved that if the defining matrix $A$ is non-singular by columns
     (abbreviated to NSC, see Definition \ref{definition3.6} for details), then $d[C(A)]\geq \min \limits_{1\leq i\leq k}\{(s+1-i)d_{i}\}$.
     Moreover, they proved that
    if $A$ is NSC and it is a column permutation of an upper triangular matrix, then $d[C(A)]=\min \limits_{1\leq i\leq k}\{(s+1-i)d_{i}\}$.

\item In \cite{O2002Note}, {\"{O}}zbudak and Stichtenoth proved that
      $d[C(A)]\geq \min \limits_{1\leq i\leq k}\{D_{i}(A)d_{i}\}$ for any full-rank defining matrix $A$, where $D_{i}(A)$ denotes
      the minimum distance of the code on $\mathbb{F}_{q}^{s}$ generated by the first $i$ rows of $A$.

\item In \cite{Hernando2009Construction}, Hernando et al. proved that $d[C(A)]=\min \limits_{1\leq i\leq k}\{D_{i}(A)d_{i}\}$
      if $C_{1}\supset C_{2}\supset\cdots\supset C_{k}$ and the defining matrix $A$ has full rank. They also deduced that
      $d[C(A)]=\min \limits_{1\leq i\leq k}\{(s+1-i)d_{i}\}$ if $C_{1}\supset C_{2}\supset\cdots\supset C_{k}$ and $A$ is NSC.
\end{itemize}

By CSS construction or Hermitian construction, we can construct new $q$-ary quantum codes of large
lengths and dimensions from the matrix-product codes that are Euclidean dual-containing or Hermitian dual-containing.
The minimum distance or minimum distance lower bound of a matrix-product code can be easily determined from the above three cases.
Therefore, we desire to know under which condition the matrix-product codes will be Euclidean dual-containing or Hermitian dual-containing.
This leads to a question: what is the Euclidean dual code (resp. Hermitian dual code) of a matrix-product code over $\mathbb{F}_{q}$
(resp. $\mathbb{F}_{q^{2}}$)? To this end, the defining matrix of a matrix-product code is usually set to be a non-singular matrix.
The solution to the question is listed as follows:

1) In \cite{Blackmore2001}, Blackmore and Norton proved that
\begin{equation*}
([C_{1},C_{2},\ldots,C_{k}]A)^{\perp}=[C_{1}^{\perp},C_{2}^{\perp},\ldots,C_{k}^{\perp}] (A^{-1})^{T}
\end{equation*}
for any non-singular matrix $A$ over $\mathbb{F}_{q}$ (see Lemma \ref{lemma3.3}).

2) In \cite{Zhang2015qc}, Zhang and Ge proved that
\begin{equation*}
([C_{1},C_{2},\ldots,C_{k}] A)^{\perp_{H}}=[C_{1}^{\perp_{H}},C_{2}^{\perp_{H}},\ldots,C_{k}^{\perp _{H}}](A^{-1})^\dagger
\end{equation*}
for any non-singular matrix $A$ over $\mathbb{F}_{q^{2}}$ (see Lemma \ref{lemma4.2}).

We have two purposes in writing this article: the first is to summarize some results of this topic obtained by the author of this article and his cooperators in  \cite{Cao2020Constructioncaowang,Cao2020New,Cao2020Constructionof}; the second is to add some new results
on quasi-orthogonal matrices (resp. quasi-unitary matrices), Euclidean dual-containing (resp. Hermitian dual-containing)
matrix-product codes and $q$-ary quantum codes derived from these matrix-product codes and CSS construction
(resp. Hermitian construction). The remainder of this article is organized as follows.
Section \ref{section2} recalls some basics on classical and quantum error-correcting codes.
Section \ref{section3} is devoted to the construction of quantum codes from Euclidean dual-containing matrix-product codes related to
quasi-orthogonal matrices and NSC quasi-orthogonal matrices.
Section \ref{section4} focuses on the construction of quantum codes from Hermitian dual-containing matrix-product codes related to
quasi-unitary matrices and NSC quasi-unitary matrices.
Section \ref{section5} makes concluding remarks of this article.

\section{Preliminaries}\label{section2}

\subsection{Classical error-correcting codes}\label{subsection2.1}
Assume $q$ is a prime power. Let $\mathbb{F}_{q}$ be the finite field with $q$ elements and $\mathbb{F}_{q}^{n}$ be
the $n$-dimensional vector space over $\mathbb{F}_{q}$. Let us recall some basics on classical error-correcting codes (see \cite{MacWilliams1997The}).

\begin{definition}\label{definition2.1}
Any $q$-ary classical error-correcting code $C$ can be viewed as a nonempty set of some vector space $\mathbb{F}_{q}^{n}$.
Any vector $\mathbf{c}=(c_{1},c_{2},\ldots,c_{n})\in C$ is called the \emph{codeword} of $C$, where $c_{i}\in \mathbb{F}_{q}$ for each $i$.
Then code $C$ is said to has \emph{length} $n$. If $C$ has $K$ codewords, then $\frac{k}{n}$ is called the
\emph{rate} or \emph{efficiency} of $C$, where $k:=\mathrm{log}_{q}K$.
\end{definition}

\begin{definition}\label{definition2.2}
Given two vectors $\mathbf{u}=(u_{1},u_{2},\ldots,u_{n})$, $\mathbf{v}=(v_{1},v_{2},\ldots,v_{n})\in \mathbb{F}_{q}^{n}$, their \emph{Hamming distance} $d_{H}(\mathbf{u},\mathbf{v})$ is defined as
\begin{equation*}
d_{H}(\mathbf{u},\mathbf{v})=\sharp\{i|1\leq i\leq n,u_{i}\neq v_{i}\}.
\end{equation*}
The \emph{Hamming weight} of $\mathbf{v}$ is defined as
\begin{equation*}
w_{H}(\mathbf{v})=\sharp\{i|1\leq i\leq n,0\neq v_{i}\in \mathbb{F}_{q}\}.
\end{equation*}
\end{definition}

\begin{definition}\label{definition2.3}
Given a $q$-ary classical error-correcting code $C$ of length $n$ and codewords number $K\geq 2$. Then, the \emph{minimum distance} of $C$ is defined as
\begin{equation*}
d(C)=\min\{d_{H}(\mathbf{c},\mathbf{c}')|\mathbf{c},\mathbf{c}'\in C,\mathbf{c}\neq \mathbf{c}'\}.
\end{equation*}
\end{definition}

If $C$ is a linear code, i.e., $C$ is a $\mathbb{F}_{q}$-vector subspace of $\mathbb{F}_{q}^{n}$, then the minimum distance of $C$ is
\begin{equation*}
d(C)=\min\{w_{H}(\mathbf{c})|\mathbf{0}\neq \mathbf{c}\in C\}.
\end{equation*}

\begin{definition}\label{definition2.4}
We denote by $(n, K, d)_{q}$ (or $[n,k,d]_{q}$) a $q$-ary classical error-correcting code of length $n$, codewords number $K$
and minimum distance $d$, where $k=\mathrm{log}_{q}K$.
\end{definition}

\begin{theorem}\label{theorem2.1}
(\!\!\cite{MacWilliams1997The}) A classical error-correcting code of minimum distance $d$ can detect up to $d-1$ errors and correct up to $\lfloor\frac{d-1}{2}\rfloor$ errors.
Here, $\lfloor x\rfloor$ denotes the greatest integer less than or equal to $x$.
\end{theorem}

\begin{theorem}\label{theorem2.2}
(\!\!\cite{MacWilliams1997The}, Singleton bound) An $[n,k,d]_{q}$ error-correcting code satisfies $d\leq n+1-k$.
\end{theorem}

\begin{definition}\label{definition2.5}
If an $[n,k,d]_{q}$ error-correcting code $C$ satisfies $d=n+1-k$, then $C$ is called a \emph{maximum distance separable}
(abbreviated to MDS) code.
\end{definition}

\subsection{GRS codes and extended GRS codes}\label{subsection2.2}
Let $k$ and $n$ be two positive integers with $k\leq n\leq q$.
Let $\mathbf{a}=(a_{1},\ldots,a_{n})$, where $a_{1},\ldots,a_{n}$ are distinct elements in $\mathbb{F}_{q}$.
Let $\mathbf{v}=(v_{1},\ldots,v_{n})$, where $v_{1},\ldots,v_{n}$ are nonzero elements in $\mathbb{F}_{q}$.
Define
\begin{align*}
GRS_{k}(\mathbf{a},\mathbf{v})=\{(v_{1}f(a_{1}),\ldots,v_{n}f(a_{n}))|f(x)\in \mathbb{F}_{q}[x], \mbox{deg}(f(x))\leq k-1\}
\end{align*}
the \emph{generalized Reed-Solomon code} (abbreviated to GRS code) associated with $\mathbf{a}$ and $\mathbf{v}$.
It is an $[n,k]_{q}$ MDS code and it has a generator matrix
$$\left[
\begin{array}{cccc}
v_{1}& v_{2}&  \cdots &v_{n}\\[4pt]
v_{1}a_{1}&v_{2}a_{2}& \cdots &v_{n}a_{n}\\
\vdots&\vdots&\ddots&\vdots\\
v_{1}a_{1}^{k-1}& v_{2}a_{2}^{k-1}& \ldots &v_{n}a_{n}^{k-1}  \\
\end{array}\right].$$
Moreover, the \emph{extended generalized Reed-Solomon code} (abbreviated to extended GRS code)
associated with $\mathbf{a}$ and $\mathbf{v}$ is defined as
$$GRS_{k}(\mathbf{a},\mathbf{v},\infty)=\{(v_{1}f(a_{1}),\ldots,v_{n}f(a_{n}),f_{k-1})|f(x)\in \mathbb{F}_{q}[x], \mbox{deg}(f(x))\leq k-1\},$$
where $f_{k-1}$ represents the coefficient of $x^{k-1}$ in $f(x)$. It is easily verified that $GRS_{k}(\mathbf{a},\mathbf{v},\infty)$ is an $[n+1,k]_{q}$ MDS code with a generator matrix
$$\left[
\begin{array}{ccccc}
v_{1}& v_{2}&  \cdots &v_{n}&0\\[4pt]
v_{1}a_{1}&v_{2}a_{2}& \cdots &v_{n}a_{n}&0\\
\vdots&\vdots&\ddots&\vdots&\vdots\\
v_{1}a_{1}^{k-2}& v_{2}a_{2}^{k-2}& \ldots &v_{n}a_{n}^{k-2}&0  \\[4pt]
v_{1}a_{1}^{k-1}& v_{2}a_{2}^{k-1}& \ldots &v_{n}a_{n}^{k-1}&1  \\
\end{array}\right].$$

\subsection{Matrix-product codes}\label{subsection2.3}

We denote by $\mathcal{M}(\mathbb{F}_{q},s\times l)$ the set of $s\times l$ matrices with entries in $\mathbb{F}_{q}$.

\begin{definition}\label{definition2.7}
(\!\!\cite{Blackmore2001}) Let $A=(a_{ij})_{i,j=1}^{s,l}\in \mathcal{M}(\mathbb{F}_{q},s\times l)$ be row full rank with $s\leq l$. Let $C_{1},C_{2},\ldots,C_{s}$
be linear codes of the same length $n$ over $\mathbb{F}_{q}$. The \emph{matrix-product code}
\begin{equation*}
C(A):=[C_{1},C_{2},\ldots,C_{s}]A,
\end{equation*}
is defined as the set of all matrix-products $[\mathbf{c}_{1},\mathbf{c}_{2},\cdots,\mathbf{c}_{s}]A$, where $A$ is called the \emph{defining matrix} of $C(A)$, and
$\mathbf{c}_{i}=(c_{1i},c_{2i},\ldots,c_{ni})^{T}\in C_{i}$ is an $n\times 1$ column vector for each $i$.
Clearly, $C(A)$ is a linear code of length $ln$.
Then, any codeword $\mathbf{c}=[\mathbf{c}_{1},\mathbf{c}_{2},\ldots,\mathbf{c}_{s}]A$ of $C(A)$ is an
$n\times l$ matrix as follows:
\begin{equation*}
\mathbf{c}=\left[
\begin{array}{cccc}
\sum_{i=1}^s c_{1i}a_{i1}& \sum_{i=1}^s c_{1i}a_{i2}& \cdots &\sum_{i=1}^s c_{1i}a_{il}\\
\sum_{i=1}^s c_{2i}a_{i1}& \sum_{i=1}^s c_{2i}a_{i2}& \cdots &\sum_{i=1}^s c_{2i}a_{il}\\
\vdots&\vdots&\ddots&\vdots\\
\sum_{i=1}^s c_{ni}a_{i1}& \sum_{i=1}^s c_{ni}a_{i2}& \cdots &\sum_{i=1}^s c_{ni}a_{il}\\
\end{array}\right].
\end{equation*}
\end{definition}

Note that $\mathbf{c}$ can be also viewed as an $1\times ln$ row vector
\begin{equation*}
\mathbf{c}=\Big[\sum_{i=1}^s a_{i1}\mathbf{c}_{i},\sum_{i=1}^s a_{i2}\mathbf{c}_{i},\ldots,\sum_{i=1}^s a_{il}\mathbf{c}_{i}\Big],
\end{equation*}
where $\mathbf{c}_{i}=(c_{1i},c_{2i},\ldots,c_{ni})\in C_{i}$ is regarded as an $1\times n$ row vector for $i=1,2,\ldots,s$.

\subsection{Quantum codes}\label{subsection2.4}

We use $\mathbb{C}^{q}$ to represent the $q$-dimensional complex vector space over the complex field $\mathbb{C}$.
For any pure $1$-qudit $|v\rangle\in \mathbb{C}^{q}$, it can be written as
$|v\rangle=\sum_{a\in \mathbb{F}_{q}} v_{a} |a\rangle$,
where $\{|a\rangle:a\in \mathbb{F}_{q}\}$ is a basis of $\mathbb{C}^{q}$, $v_{a}\in \mathbb{C}$ and
$\sum_{a\in \mathbb{F}_{q}} |v_{a}|^{2}=1$.
Any $n$-qudit is a joint state of $n$ qudits of the $q^{n}$-dimensional complex vector space
$(\mathbb{C}^{q})^{\otimes n}\cong\mathbb{C}^{q^{n}}$. We can write any pure $n$-qudit as
$|\mathbf{v}\rangle=\sum_{\mathbf{a}\in \mathbb{F}_{q}^{n}} v_{\mathbf{a}} |\mathbf{a}\rangle$,
where
$\{|\mathbf{a}\rangle=|a_{1}\rangle \otimes \cdots \otimes|a_{n}\rangle:(a_{1},\ldots,a_{n})\in \mathbb{F}_{q}^{n}\}$
is a basis of $\mathbb{C}^{q^{n}}$, $v_{\mathbf{a}}\in \mathbb{C}$ and
$\sum_{\mathbf{a}\in \mathbb{F}_{q}^{n}} |v_{\mathbf{a}}|^{2}=1$.

Let $\gamma$ be a complex primitive $p$-th root of unity.
Let $\mathbf{a}=(a_{1},\ldots,a_{n})$, $\mathbf{b}=(b_{1},\ldots,b_{n})\in \mathbb{F}_{q}^{n}$.
Define the error operators $T(a_{i})$ and $R(a_{i})$ as
$T(a_{i})|x\rangle=|x+a_{i}\rangle$ and $R(a_{i})|x\rangle=\gamma^{\mathrm{Tr}(a_{i}x)}|x\rangle$, respectively, where
$\mathrm{Tr}(x)$ is the trace function from $\mathbb{F}_{q}$ ($q=p^{m}$ for some $m\in \mathbb{N}^{+}$) to $\mathbb{F}_{p}$.

If we use $T(\mathbf{a})=T(a_{1})\otimes\cdots\otimes T(a_{n})$ and $R(\mathbf{a})=R(a_{1})\otimes\cdots\otimes R(a_{n})$
to represent the tensor products of $n$ error operators, then the error set
$$E_{n}=\{\gamma^{i}T(\mathbf{a})R(\mathbf{b})|0\leq i\leq p-1,\mathbf{a},\mathbf{b}\in \mathbb{F}_{q}^{n}\}$$
forms an error group.
For any error $\mathbf{e}=\gamma^{i}T(\mathbf{a})R(\mathbf{b})\in E_{n}$, its \emph{quantum weight} is defined as
$$w_{Q}(\mathbf{e})=\sharp\{i|(a_{i},b_{i})\neq (0,0)\}.$$

Let $E_{n}(i)=\{\mathbf{e}\in E_{n}|w_{Q}(\mathbf{e})\leq i\}$.
For a $q$-ary quantum code $Q$, if $d$
is the largest positive integer such that $\langle \mathbf{u}|\mathbf{e}|\mathbf{v}\rangle=\mathbf{0}$ holds for any
$|\mathbf{u}\rangle,|\mathbf{v}\rangle\in Q$ with
$\langle \mathbf{u}|\mathbf{v}\rangle=\mathbf{0}$ and
$\mathbf{e}\in E_{n}(d-1)$, then we say $Q$ has \emph{minimum distance} $d$.
Usually, we use the notation $[[n,k,d]]_{q}$ to represent a $q$-ary quantum code of length $n$, dimension $q^{k}$ and minimum distance $d$.
It has the abilities to detect up to $d-1$ quantum errors and correct up to $\lfloor \frac{d-1}{2}\rfloor$ quantum errors.
The minimum distance $d$ of a quantum code must satisfy the \emph{quantum Singleton bound}, i.e., $2d\leq n+2-k$.
Further, if $2d=n+2-k$, then such a quantum code is called a \emph{quantum MDS code}.
For more information on quantum codes, see \cite{Shor1995Scheme,Calderbank1998Quantumerror,Steane1996Errorcorrecting,Calderbank1996Goodq,Steane1996Simpleq,
Calderbank1997Quantumerrorcorrection,Rains1999Nonbinary,Ashikhmin2001,Ketkar2006Nonbinary}.

\section{Quantum codes from Euclidean dual-containing matrix-product codes}\label{section3}

\subsection{Basic concepts and properties}\label{subsection3.1}
Assume $q$ is a prime power. Let $\mathbb{F}_{q}$ be the finite field with $q$ elements and let $\mathbb{F}_{q}^{\ast}$ denote the set of non-zero elements over $\mathbb{F}_{q}$.

\begin{definition}\label{definition3.1}
For
$\mathbf{x}=(x_{1},x_{2},\ldots,x_{n})$, $\mathbf{y}=(y_{1},y_{2},\ldots,y_{n})$ in $\mathbb{F}_{q}^{n}$, define
\begin{equation*}
(\mathbf{x},\mathbf{y})=\sum_{i=1}^n x_i y_i
\end{equation*}
as the \emph{Euclidean inner product} of $\mathbf{x}$ and $\mathbf{y}$.
For a liner code $C$ of length $n$ over $\mathbb{F}_{q}$, define
\begin{equation*}
C^\perp=\{\mathbf{x}\in\mathbb{F}_{q}^{n}|(\mathbf{x},\mathbf{y})=0 \ \mathrm{for} \ \mathrm{all} \  \mathbf{y}\in C\}
\end{equation*}
as the \emph{Euclidean dual code} of $C$.
\end{definition}

\begin{definition}\label{definition3.2}
Let $C$ be a linear code over $\mathbb{F}_{q}^{n}$. Then,

(1) If $C\subseteq C^{\perp}$, then $C$ is called an \emph{Euclidean self-orthogonal code};

(2) If $C^{\perp}=C$, then $C$ is called an \emph{Euclidean self-dual code};

(3) If $C^{\perp}\subseteq C$, then $C$ is called an \emph{Euclidean dual-containing code}.

\end{definition}

The following CSS construction
tells us how to produce $q$-ary quantum codes from
the classical linear codes over $\mathbb{F}_{q}$.

\begin{theorem} \label{theorem3.1}
(CSS construction, \cite{Calderbank1996Goodq,Steane1996Simpleq}) Let $C$ be an $[n,k,d]_{q}$ linear code
with $C^\perp \subseteq C$, then there exists an $[[n,2k-n,\geq d]]_{q} $ quantum code.
\end{theorem}

The following lemma characterizes the parameters of the matrix-product codes over $\mathbb{F}_{q}$.

\begin{lemma}\label{lemma3.1}
(\!\!\cite{O2002Note}) Let $C_{i}$ be an $[n,t_{i},d_{i}]_{q}$ linear code for $i=1,2,\ldots,k$. Let $A\in
\mathcal{M}(\mathbb{F}_{q},k\times s)$ be full-rank.
Denote by $D_{i}(A)$ the minimum distance of the code on $\mathbb{F}_{q}^{s}$ generated by the first $i$ rows of $A$.
Then, the matrix-product code
\begin{equation*}
C(A)=[C_{1},C_{2},\ldots,C_{k}]A
\end{equation*}
is an $[sn,\sum_{i=1}^k t_i,\geq d]_{q}$ linear code, where
$d=\min \limits_{1\leq i\leq k}\{D_{i}(A)d_{i}\}$.
\end{lemma}

The following lemma gives the Euclidean dual code of a matrix-product code over $\mathbb{F}_{q}$.

\begin{lemma}\label{lemma3.3}
(\!\!\cite{Blackmore2001}) Let $C_{i}$ be an $[n,t_{i},d_{i}]_{q}$ linear code, where $i=1,2,\ldots,k$. Let $A\in
M(\mathbb{F}_{q},k\times k)$ be non-singular. Then,
\begin{equation*}
([C_{1},C_{2},\ldots,C_{k}] A)^{\perp}=[C_{1}^{\perp},C_{2}^{\perp},\ldots,C_{k}^{\perp}] (A^{-1})^{T}.
\end{equation*}
\end{lemma}

\subsection{General approach for constructing quantum codes via Euclidean dual-containing matrix-product codes}\label{subsection3.2}

Let $$\tau=\left(
\begin{array}{cccc}
1& 2&  \cdots &k\\
i_{1}& i_{2}& \cdots &i_{k}  \\
\end{array}\right)$$
denote a permutation on $ \{1,2,\ldots,k\}$. Let $P_{\tau}$
be the $k\times k$ permutation matrix in which the $i_{j}$-th row
of the identity matrix $I_{k}$ is replaced with the $j$-th row of it for $j=1,2,\ldots,k$.

\begin{definition}\label{definition3.3}
Let $B\in \mathcal{M}(\mathbb{F}_{q},k\times k)$. If $B=DP_{\tau}$, where $D=\mathrm{diag}(d_{11},\ldots,d_{kk})$ with $d_{ii} \in \mathbb{F}_{q}^{\ast}$ for each $i$, then the matrix $B$ is called a \emph{monomial matrix} with respect to the permutation $\tau$.
\end{definition}

\begin{definition}\label{definition3.4}
Let $B\in \mathcal{M}(\mathbb{F}_{q},k\times k)$. If $BB^{T}$ is diagonal over $\mathbb{F}_{q}^{\ast}$,
then we call $B$ a \emph{quasi-orthogonal matrix} over $\mathbb{F}_{q}$.
If $BB^{T}=I_{k}$, then we call $B$ an \emph{orthogonal matrix} over $\mathbb{F}_{q}$.
\end{definition}

The following theorem gives a general approach for constructing $q$-ary quantum codes via Euclidean dual-containing matrix-product
codes over $\mathbb{F}_{q}$.

\begin{theorem}\label{theorem3.2}
(\!\!\cite{Cao2020New}) Let $C_{j}$ be an $[n,t_{j},d_{j}]_{q}$ linear code with $C_{i_{j}}^{\perp}\subseteq C_{j}$, where $j=1,2,\ldots,k$.
For any non-singular matrix $A\in \mathcal{M}(\mathbb{F}_{q},k\times k)$,
if $AA^{T}$ is a monomial matrix with respect to the permutation $\tau$, then the matrix-product code
\begin{equation*}
C(A)=[C_{1},C_{2},\ldots,C_{k}]A
\end{equation*}
is an $[kn,\sum_{i=1}^k t_i,\geq d]_{q}$ Euclidean dual-containing code, where $d=\min \limits_{1\leq i\leq k}\{d_{i}D_{i}(A)\}$.
Further, $C(A)$ generates an $[[kn,2\sum_{i=1}^k t_i-kn,\geq d]]_{q}$ quantum code.
\end{theorem}

By Theorem \ref{theorem3.2}, we can construct Euclidean dual-containing matrix-product code $C(A)$ over $\mathbb{F}_{q}$,
as long as the following two conditions hold:

(a) $AA^{T}$ is a monomial matrix with respect to $\tau$, where $\tau$ maps each $j\in \{1,2,\ldots,k\}$ to $i_{j}\in \{1,2,\ldots,k\}$;

(b) The constituent codes $C_{1},C_{2},\ldots,C_{k}$ of $C(A)$ satisfy $C_{i_{j}}^{\perp}\subseteq C_{j}$ for each $j=1,2,\ldots,k$.

Further, we can obtain $q$-ary quantum codes by the CSS construction. In fact, for a given permutation $\tau$, it is not difficult to find proper constituent codes satisfying (b). We emphasize that the main difficulty lies in condition (a).
As far as we know, a general method for finding the matrix $A$ in condition (a) is still lacking at present.
In the next subsections, we will investigate the constructions of quantum codes when the defining matrix
$A$ of $C(A)$ in turn satisfies different cases: (i) $A$ is a quasi-orthogonal matrix; (ii) $A$ is a NSC quasi-orthogonal matrix.

\subsection{Quantum codes related to quasi-orthogonal matrices}\label{subsection3.3}

When $\tau$ is an identity permutation, i.e., $\tau=(1)$, it follows from Theorem \ref{theorem3.2} that

\begin{corollary}\label{corollary3.1}
Let $C_{i}$ be an $[n,t_{i},d_{i}]_{q}$ linear code with $C_{i}^{\perp}\subseteq C_{i}$ for $i=1,2,\ldots,k$.
For any non-singular matrix $A\in \mathcal{M}(\mathbb{F}_{q},k\times k)$,
if $A$ is quasi-orthogonal, then the matrix-product code
\begin{equation*}
C(A)=[C_{1},C_{2},\ldots,C_{k}]A
\end{equation*}
is an $[kn,\sum_{i=1}^k t_i,\geq d]_{q}$ Euclidean dual-containing code, where $d=\min \limits_{1\leq i\leq k}\{d_{i}D_{i}(A)\}$.
Further, $C(A)$ generates an $[[kn,2\sum_{i=1}^k t_i-kn,\geq d]]_{q}$ quantum code.
\end{corollary}

By Corollary \ref{corollary3.1}, the next task is to find the quasi-orthogonal matrix $A$ for constructing quantum codes.

\subsubsection{The utilization of the theory of quadratic forms}\label{subsubsection3.3.1}

Similar to the theory of real quadratic forms,
any quadratic form in $\mathbb{F}_{q}$, where $q$ is an odd prime power, can be simplified into a diagonal
form according to a linear invertible transform. In other words, this property is equivalent to
the following proposition, which is shown by the language of matrices. The reader can refer \cite{Birkhoff1998A} for more information.

\begin{proposition}\label{proposition3.1}
(\!\!\cite{Birkhoff1998A}) Let $q$ be an odd prime power. Then for any symmetric matrix
$A\in \mathcal{M}(\mathbb{F}_{q},k\times k)$, there exists a non-singular matrix
$M\in \mathcal{M}(\mathbb{F}_{q},k\times k)$ such that $M^{T}AM$
is a $k\times k$ diagonal matrix.
\end{proposition}

By the above proposition, the following theorem constructs Euclidean dual-containing matrix-product codes
and obtains quantum codes by CSS construction.

\begin{theorem}\label{theorem3.3}
(\!\!\cite{Cao2020New}) Let $q$ be an odd prime power and $C_{i}$ be an $[n,t_{i},d_{i}]_{q}$ linear code with $C_{i}^{\perp}\subseteq C_{i}$ for $i=1,2,\ldots,k$. Then, for any non-singular matrix $B\in \mathcal{M}(\mathbb{F}_{q},k\times k)$, there exists a non-singular matrix $N\in \mathcal{M}(\mathbb{F}_{q},k\times k)$ such that the matrix-product code
\begin{equation*}
C(N^{T}B)=[C_{1},C_{2},\ldots,C_{k}]N^{T}B
\end{equation*}
is an $[kn,\sum_{i=1}^k t_i,\geq d]_{q}$ Euclidean dual-containing code, where $d=\min \limits_{1\leq i\leq k}\{d_{i}{D_{i}(N^{T}B)}\}$.
Further, $C(N^{T}B)$ yields an $[[kn,2\sum_{i=1}^k t_i-kn,\geq d]]_{q}$ quantum code.
\end{theorem}

\subsubsection{The utilization of the theory of quadratic sum}\label{subsubsection3.3.2}

The following proposition gives an interesting property on the quadratic sum over finite fields.

\begin{proposition}\label{proposition3.2}
(\!\!\cite{Feng2011Quadratic}) Let $\mathbb{F}$ be a field and let $c=c_{1}^{2}+\cdots+c_{2^{m}}^{2}$ with $c_{i}\in \mathbb{F}$ for each $i$.
Then, there exist $4^{m}$ elements $\{s_{i,j}|1\leq i,j\leq 2^{m}\}$ over $\mathbb{F}$, satisfying that

(1) $s_{1,j}=c_{j},1\leq j\leq 2^{m}$;

(2) $\sum_{k=1}^{2^{m}} s_{i,k}s_{j,k}=\sum_{k=1}^{2^{m}} s_{k,i}s_{k,j}=\delta_{i,j}c$, where $\delta_{i,j}$ is the Kronecker symbol
for $1\leq i,j\leq 2^{m}$.
\end{proposition}

Proposition \ref{proposition3.2} is equivalent to the following proposition through the form of matrix language.

\begin{proposition}\label{proposition3.3}
(\!\!\cite{Feng2011Quadratic}) Let $\mathbb{F}$ be a field and let $c=c_{1}^{2}+\cdots+c_{2^{m}}^{2}$ with $c_{i}\in \mathbb{F}$ for each $i$.
Then, there exists a matrix $S=(s_{i,j})\in \mathcal{M}(\mathbb{F},2^{m}\times 2^{m})$ with $(c_{1},\ldots,c_{2^{m}})$ being the first row of $A$, such that
\begin{equation*}
S^{T}S=SS^{T}=cI_{2^{m}}.
\end{equation*}
\end{proposition}

By Proposition \ref{proposition3.3}, we have the following theorem.

\begin{theorem}\label{theorem3.4}
Let $0\neq c=c_{1}^{2}+\cdots+c_{2^{m}}^{2}$, where $c_{i}\in \mathbb{F}_{q}$ for $i=1,2,\ldots,2^{m}$.
Let $C_{i}$ be an $[n,t_{i},d_{i}]_{q}$ linear code with $C_{i}^{\bot}\subseteq C_{i}$ for $i=1,2,\ldots,2^{m}$.
Then, there exists a matrix $S=(s_{i,j})\in \mathcal{M}(\mathbb{F}_{q},2^{m}\times 2^{m})$ such that the matrix-product code
\begin{equation*}
C(S)=[C_{1},C_{2},\ldots,C_{2^{m}}]S
\end{equation*}
is an $[2^{m}n,\sum_{i=1}^{2^{m}} t_i,\geq d]_{q}$ Euclidean dual-containing code,
where $d=\min \limits_{1\leq i\leq 2^{m}}\{d_{i}D_{i}(S)\}$.
Further, $C(S)$ generates an $[[2^{m}n,2\sum_{i=1}^{2^{m}} t_i-2^{m}n,\geq d]]_{q}$ quantum code.
\end{theorem}

\subsubsection{The utilization of the Hadamard matrices}\label{subsubsection3.3.3}

\begin{definition}\label{definition3.5}
Assume $H=(h_{ij})$ is an $n\times n$ matrix with $h_{ij}=\pm1$. If $H_{n}H_{n}^{T}=nI_{n}$,
then we call $H_{n}$ a \emph{Hadamard matrix}.
\end{definition}

It is not difficult to verify that the following three matrices
\begin{equation*}
H_{1}=[1],
H_{2}=\left[
\begin{array}{cc}
1&1\\
1&-1\\
\end{array}\right],
H_{4}=\left[
\begin{array}{cccc}
1&1&1&1\\
1&-1&1&-1\\
1&1&-1&-1\\
1&-1&-1&1\\
\end{array}\right]
\end{equation*}
are all Hadamard matrices. The following proposition gives a class of Hadamard matrices over $\mathbb{F}_{q}$.

\begin{proposition}\label{proposition3.4}
(\!\!\cite{Lidl1997Finite}) Let $\mathbb{F}_{q}=\{a_{1},a_{2},\ldots,a_{q}\}$, $q\equiv3\mod{4}$. Denote by $\eta$ the quadratic character of $\mathbb{F}_{q}$. Then,
\begin{equation*}
H=\left[
\begin{array}{cccccc}
1&1&1&1&\cdots&1\\
1&-1&b_{12}&b_{13}&\cdots&b_{1q}\\
1&b_{21}&-1&b_{23}&\cdots&b_{2q}\\
1&b_{31}&b_{32}&-1&\cdots&b_{3q}\\
\vdots&\vdots&\vdots&\vdots&\ddots&\vdots\\
1&b_{q1}&b_{q2}&b_{q3}&\cdots&-1\\
\end{array}\right]
\end{equation*}
is an $(q+1)\times (q+1)$ Hadamard matrix, where $b_{ij}=\eta(a_{j}-a_{i})$ for $1\leq i\neq j\leq q$.
\end{proposition}

\begin{corollary}\label{corollary3.2}
(\!\!\cite{Lidl1997Finite}) Assume $q\equiv3\mod{4}$ and $H$ is defined as in Proposition \ref{proposition3.4}.
Define $H_{0}=H$,
\begin{equation*}
H_{w}:=\left[
\begin{array}{cc}
H_{w-1}&H_{w-1}\\
H_{w-1}&-H_{w-1}\\
\end{array}\right],w\geq 1.
\end{equation*}
Then, $H_{w}$ is an $2^{w}(q+1)\times 2^{w}(q+1)$ Hadamard matrix for any integer $w$.
\end{corollary}

By Corollary \ref{corollary3.2}, we immediately construct Euclidean dual-containing matrix-product codes and obtain
the corresponding quantum codes in the following theorem.

\begin{theorem}\label{theorem3.5}
Assume $q\equiv3\mod{4}$, $w\geq 1$ and $H_{w}$ is defined as in Corollary \ref{corollary3.2}. Write $w'=2^{w}(q+1)$.
Let $C_{i}$ be an $[n,t_{i},d_{i}]_{q}$ linear code with $C_{i}^{\bot}\subseteq C_{i}$ for each $i=1,\ldots,w'$.
Then, the matrix-product code
\begin{equation*}
C(H_{w})=[C_{1},C_{2},\ldots,C_{w'}]H_{w}
\end{equation*}
is an $[w'n,\sum_{i=1}^{w'} t_i,\geq d]_{q}$ Euclidean dual-containing code,
where $d=\min \limits_{1\leq i\leq w'}\{d_{i}D_{i}(H_{w})\}$. Further, $C(H_{w})$ generates an $[[w'n,2\sum_{i=1}^{w'} t_i-w'n,\geq d]]_{q}$
quantum code.
\end{theorem}

\subsection{Quantum codes related to NSC quasi-orthogonal matrices}\label{subsection3.4}

Recall that in Corollary \ref{corollary3.1} when the defining matrix $A$ is quasi-orthogonal, we can construct an
Euclidean dual-containing matrix-product code.
Further, it generates an $[[kn,2\sum_{i=1}^k t_i-kn,\geq d]]_{q}$ quantum code from the CSS construction, where
$d=\min \limits_{1\leq i\leq k}\{d_{i}D_{i}(A)\}$. Clearly, the length and dimension of the quantum code are irrelevant to the defining matrix $A$,
while the minimum distance is related to it and it is determined by $D_{i}(A)$. Moreover, it is not difficult to check that $D_{i}(A)\leq k+1-i$.

When fixing the length and dimension, the larger value of the minimum distance means the better performance of error detection and error correction of the
quantum code. Given this fact, we wish to make the minimum distance lower bound $\min \limits_{1\leq i\leq k}\{d_{i}D_{i}(A)\}$ of the quantum codes from our constructed matrix-product codes as large as possible.
That is to say, we need to find the quasi-orthogonal matrix $A$ such that $D_{i}(A)=k+1-i$.
For general quasi-orthogonal matrix $A$, the computation of $D_{i}(A)$ is more and more difficult as the order of $A$ increases.

\begin{definition}\label{definition3.6}
(\!\!\cite{Blackmore2001}) Let $A=(a_{ij})\in \mathcal{M}(\mathbb{F}_{q},k\times k)$.
Denote by $D_{i}(A)$ the minimum distance of the code on $\mathbb{F}_{q}^{k}$ generated by the first $i$ rows of $A$.
Write $A^{(i)}$ the matrix consisting of the first $i$ rows of $A$ and
$A(j_{1},\ldots,j_{i})$ the matrix consisting of the $j_{1},\ldots,j_{i}$ columns of $A^{(i)}$, where $1\leq j_{1}<\ldots< j_{i}\leq k$.
If $A(j_{1},\ldots,j_{i})$ is non-singular for all $1\leq i\leq k$ and $1\leq j_{1}<\ldots< j_{i}\leq k$,
then we call $A$ \emph{non-singular by columns (NSC)}.
\end{definition}

From Definition \ref{definition3.6}, one can verify that a NSC matrix $A$ satisfies $D_{i}(A)=k+1-i$ exactly.
If $A$ is both NSC and quasi-orthogonal, then
we call $A$ \emph{NSC quasi-orthogonal}. Next, we will give a constructive method for acquiring general quasi-orthogonal matrices
and NSC quasi-orthogonal matrices.

\begin{theorem}\label{theorem3.6}
Let $A\in\mathcal{M}(\mathbb{F}_{q},k\times k)$ be non-singular. If all leading principal minors of $AA^{T}$ are nonzero,
then there exists a lower unitriangular matrix $L$ such that $LA$ is quasi-orthogonal over $\mathbb{F}_{q}$.
Further, if $A$ is NSC, then $LA$ is NSC quasi-orthogonal.
\end{theorem}

\begin{proof}
Suppose
\begin{equation*}
AA^{T}=\left[
\begin{array}{cc}
A_{k-1}&\mathbf{h}\\[6pt]
\mathbf{h}^{T}&c\\
\end{array}\right],
\end{equation*}
where $A_{k-1}=A_{k-1}^{T}$. Let
\begin{equation*}
L_{k-1}=\left[
\begin{array}{cc}
I_{k-1}&\mathbf{0}_{(k-1)\times1}\\
-\mathbf{h}^{T}A_{k-1}^{-1}&1\\
\end{array}\right],
\end{equation*}
then 
\begin{equation*}
L_{k-1}AA^{T}L_{k-1}^{T}=\left[
\begin{array}{cc}
A_{k-1}&\mathbf{0}_{(k-1)\times1}\\
\mathbf{0}_{1\times(k-1)}&c-\mathbf{h}^{T}A_{k-1}^{-1}\mathbf{h}\\
\end{array}\right].
\end{equation*}

Repeating such process enables us to get a lower unitriangular matrix $L$ such that
\begin{equation*}
LAA^{T}L^{T}=\mbox{diag}(\lambda_{1},\lambda_{2},\ldots,\lambda_{k}),
\end{equation*}
where $\lambda_{i}\neq 0$ for each $i$. That is to say, $LA$ is quasi-orthogonal.

Let us now prove that $LA$ is NSC if $A$ is NSC. Let $A=(a_{ij})_{i,j=1}^{k}$ and $L=(l_{ij})_{i,j=1}^{k}$. For any $1\leq s\leq k$ and
$1\leq j_{1}<\cdots< j_{s}\leq k$, it is verified that
\begin{align*}
|(LA)(j_{1},\ldots,j_{s})|&=\left|
\begin{array}{cccc}
a_{1j_{1}}&a_{1j_{2}}&\cdots&a_{1j_{s}}\\
\sum_{i=1}^2 l_{2i} a_{ij_{1}}&\sum_{i=1}^2 l_{2i} a_{ij_{2}}&\cdots&\sum_{i=1}^2 l_{2i} a_{ij_{s}}\\
\vdots&\vdots&\ddots&\vdots\\
\sum_{i=1}^s l_{si} a_{ij_{1}}&\sum_{i=1}^s l_{si} a_{ij_{2}}&\cdots&\sum_{i=1}^s l_{si} a_{ij_{s}}\\
\end{array}\right|=\left|
\begin{array}{cccc}
a_{1j_{1}}&a_{1j_{2}}&\cdots&a_{1j_{s}}\\
a_{2j_{1}}&a_{2j_{2}}&\cdots&a_{2j_{s}}\\
\vdots&\vdots&\ddots&\vdots\\[6pt]
a_{sj_{1}}&a_{sj_{2}}&\cdots&a_{sj_{s}}\\
\end{array}\right|\\
& =|A(j_{1},\ldots,j_{s})|.
\end{align*}
When $A$ is NSC, we have $|(LA)(j_{1},\ldots,j_{s})|=|A(j_{1},\ldots,j_{s})|\neq 0$.
By Definition \ref{definition3.6}, $LA$ is NSC as well. Therefore, $LA$ is NSC quasi-orthogonal. $\hfill\square$
\end{proof}

\begin{example}\label{example3.2}
In $\mathbb{F}_{5}$, let us consider the NSC matrix
$
A=\left[
\begin{array}{ccc}
1&1&2\\
2&0&3\\
1&4&0\\
\end{array}\right]$. Then 
$AA^{T}=\left[
\begin{array}{ccc}
1&3&0\\
3&3&2\\
0&2&2\\
\end{array}\right]$. It is easily verified that all leading principal minors of $AA^{T}$ are nonzero. By Theorem \ref{theorem3.6}, there exists
\begin{align*}
L_{2}=\left[
\begin{array}{cc|c}
1&0&0\\
0&1&0\\\hline
4&2&1\\
\end{array}\right]
\end{align*}
such that
\begin{align*}
L_{2}AA^{T}L_{2}^{T}=\left[
\begin{array}{cc|c}
1&3&0\\
3&3&0\\\hline
0&0&1\\
\end{array}\right]\triangleq\left[
\begin{array}{cc}
A_{2}&\mathbf{0}_{2\times1}\\
\mathbf{0}_{1\times2}&1\\
\end{array}\right].
\end{align*}
Then, there exists
$B_{1}=\left[
\begin{array}{cc}
1&0\\
2&1\\
\end{array}\right]$
such that
$B_{1}A_{2}B_{1}^{T}=\left[
\begin{array}{cc}
1&0\\
0&4\\
\end{array}\right]$. Let
$L_{1}=\left[
\begin{array}{cc}
B_{1}&\mathbf{0}_{2\times1}\\
\mathbf{0}_{1\times2}&1\\
\end{array}\right]$, then we obtain
\begin{align*}
L_{1}L_{2}AA^{T}L_{2}^{T}L_{1}^{T}=\left[
\begin{array}{ccc}
1&0&0\\
0&4&0\\
0&0&1\\
\end{array}\right].
\end{align*}
Let $L=L_{1}L_{2}$, then we obtain a lower unitriangular matrix $L$ such that $LAA^{T}L^{T}=\mathrm{diag}(1,4,1)$.
Hence, $LA$ is quasi-orthogonal over $\mathbb{F}_{5}$. Moreover, it is not difficult to find that
\begin{align*}
LA=\left[
\begin{array}{ccc}
1&1&2\\
4&2&2\\
4&3&4\\
\end{array}\right]
\end{align*}
is NSC over $\mathbb{F}_{5}$. Therefore, $LA$ is NSC quasi-orthogonal over $\mathbb{F}_{5}$.
\end{example}

By Theorem \ref{theorem3.6}, we obtain the following result.

\begin{theorem}\label{theorem3.7}
Let $C_{i}$ be an $[n,t_{i},d_{i}]_{q}$ linear code with $C_{i}^{\perp}\subseteq C_{i}$ for each $i=1,2,\ldots,k$.
For any NSC matrix $A\in\mathcal{M}(\mathbb{F}_{q},k\times k)$, if all leading principal minors of $AA^{T}$ are nonzero,
then there exists a lower unitriangular matrix $L$ such that the matrix-product code
\begin{equation*}
C(LA)=[C_{1},C_{1},\ldots,C_{k}]LA
\end{equation*}
is an $[kn,\sum_{i=1}^k t_i,\geq d]_{q}$ Euclidean dual-containing code,
where $d=\min \limits_{1\leq i\leq k}\{(k+1-i)d_{i}\}$.
Further, $C(LA)$ generates an $[[kn,2\sum_{i=1}^k t_i-kn,\geq d]]_{q}$ quantum code.
\end{theorem}

\begin{proof}
By Theorem \ref{theorem3.6}, there exists a lower unitriangular matrix $L\in\mathcal{M}(\mathbb{F}_{q},k\times k)$ such that
$LAA^{T}L^{T}=R$, where $R=\mbox {diag}(r_{11},\ldots,r_{kk})$ with each $r_{ii}\in \mathbb{F}_{q}^{\ast}$.
Then, we have $[(LA)^{-1}]^{T}=R^{-1}LA$. By Lemma \ref{lemma3.3} we know
\begin{equation*}
([C_{1},C_{2},\ldots,C_{k}]LA)^{\perp}=[C_{1}^{\perp},C_{2}^{\perp},\ldots,C_{k}^{\perp}] R^{-1}LA.
\end{equation*}
Besides, it follows from $r_{ii}^{-1}C_{i}^{\perp}=C_{i}^{\perp}$ that
\begin{align*}
[C_{1}^{\perp},C_{2}^{\perp},\ldots,C_{k}^{\perp}] R^{-1}LA&=[r_{11}^{-1}C_{1}^{\perp},r_{22}^{-1}C_{2}^{\perp},\ldots,r_{kk}^{-1}C_{k}^{\perp}] LA\\
&=[C_{1}^{\perp},C_{2}^{\perp},\ldots,C_{k}^{\perp}] LA\\
& \subseteq[C_{1},C_{2},\ldots,C_{k}]LA.
\end{align*}
Hence, $C(LA)$ is Euclidean dual-containing with parameters $[kn,\sum_{i=1}^k t_i,\geq d]_{q^{2}}$,
where $d=\min \limits_{1\leq i\leq k}\{(k+1-i)d_{i}\}$. By CSS construction, it generates an
$[[kn,2\sum_{i=1}^k t_i-kn,\geq d]]_{q}$ quantum code. $\hfill\square$
\end{proof}

Similar to Example \ref{example3.2}, we are able to construct the $2\times 2$ NSC quasi-orthogonal matrix over $\mathbb{F}_{5}$.
Let us consider the NSC matrix
$
\widetilde{A_{1}}=\left[
\begin{array}{cc}
2&3\\
1&2\\
\end{array}\right]$. We have
$\widetilde{A_{1}}\widetilde{A_{1}}^{T}=\left[
\begin{array}{cc}
3&3\\
3&0\\
\end{array}\right]$ whose all leading principal minors are nonzero. By Theorem \ref{theorem3.6}, there exists a lower unitriangular matrix
$\widetilde{L_{1}}=\left[
\begin{array}{cc}
1&0\\
4&1\\
\end{array}\right]$
such that
$
\widetilde{L_{1}}\widetilde{A_{1}}\widetilde{A_{1}}^{T}\widetilde{L_{1}}^{T}=\left[
\begin{array}{cc}
3&0\\
0&2\\
\end{array}\right]$. Thus, $\widetilde{L_{1}}\widetilde{A_{1}}$ is NSC quasi-orthogonal.

For convenience, let us replace the $3\times 3$ matrices $L$ and $A$ in Example \ref{example3.2}
with $\widetilde{L_{2}}$ and $\widetilde{A_{2}}$, respectively. From Theorem \ref{theorem3.7}, we have the following proposition.

\begin{proposition}\label{proposition3.5}
Let $C_{i}$ be an $[n,t_{i},d_{i}]_{5}$ linear code with $C_{i}^{\perp}\subseteq C_{i}$, $i=1,2,3$. Then,

(1) The matrix-product code $C(\widetilde{L_{1}}\widetilde{A_{1}})=[C_{1},C_{2}] \widetilde{L_{1}}\widetilde{A_{1}}$ is an
$[2n,t_{1}+t_{2},\geq d]_{5}$ Euclidean dual-containing code, where $d=\min \{2d_{1},d_{2}\}$. Further, $C(\widetilde{L_{1}}\widetilde{A_{1}})$
generates an $[[2n,2(t_{1}+t_{2}-n),\geq d]]_{5}$ quantum code;

(2) The matrix-product code $C(\widetilde{L_{2}}\widetilde{A_{2}})=[C_{1},C_{2},C_{3}] \widetilde{L_{2}}\widetilde{A_{2}}$ is an
$[3n,t_{1}+t_{2}+t_{3},\geq d]_{5}$ Euclidean dual-containing code, where $d=\min \{3d_{1},2d_{2},d_{3}\}$. Further, $C(\widetilde{L_{2}}\widetilde{A_{2}})$ generates an $[[3n,2(t_{1}+t_{2}+t_{3})-3n,\geq d]]_{5}$ quantum code.
\end{proposition}

Next, let us construct the $2\times 2$, $3\times 3$ and $4\times 4$ NSC quasi-orthogonal matrix over $\mathbb{F}_{7}$. Denote
\begin{equation*}
\widetilde{A_{3}}=\left[
\begin{array}{cc}
1&2\\
2&3\\
\end{array}\right],
\widetilde{A_{4}}=\left[
\begin{array}{ccc}
1&3&4\\
0&1&2\\
2&3&5\\
\end{array}\right],
\widetilde{A_{5}}=\left[
\begin{array}{cccc}
1&2&3&4\\
2&3&4&1\\
3&4&1&2\\
4&1&2&3\\
\end{array}\right].
\end{equation*}
One can see that $\widetilde{A_{3}}$, $\widetilde{A_{4}}$ and $\widetilde{A_{5}}$ are all NSC, and
all leading principal minors of
$\widetilde{A_{3}}\widetilde{A_{3}}^{T}$,
$\widetilde{A_{4}}\widetilde{A_{4}}^{T}$ and $\widetilde{A_{5}}\widetilde{A_{5}}^{T}$ are nonzero.
Then, it follows from Theorem \ref{theorem3.6} that there exist lower unitriangular matrices
\begin{equation*}
\widetilde{L_{3}}=\left[
\begin{array}{cc}
1&0\\
4&1\\
\end{array}\right],
\widetilde{L_{4}}=\left[
\begin{array}{ccc}
1&0&0\\
2&1&0\\
1&5&1\\
\end{array}\right],
\widetilde{L_{5}}=\left[
\begin{array}{cccc}
1&0&0&0\\
2&1&0&0\\
0&2&1&0\\
2&4&2&1\\
\end{array}\right]
\end{equation*}
such that $\widetilde{L_{3}}\widetilde{A_{3}}\widetilde{A_{3}}^{T}\widetilde{L_{3}}^{T}=\mathrm {diag}(5,3)$,
$\widetilde{L_{4}}\widetilde{A_{4}}\widetilde{A_{4}}^{T}\widetilde{L_{4}}^{T}=\mathrm {diag}(5,6,1)$,
$\widetilde{L_{5}}\widetilde{A_{5}}\widetilde{A_{5}}^{T}\widetilde{L_{5}}^{T}=\mathrm {diag}(2,1,1,4)$. Hence
\begin{align*}
\widetilde{L_{3}}\widetilde{A_{3}}=\left[
\begin{array}{cc}
1&2\\
6&4\\
\end{array}\right],
\widetilde{L_{4}}\widetilde{A_{4}}=\left[
\begin{array}{ccc}
1&3&4\\
2&0&3\\
3&4&5\\
\end{array}\right],
\widetilde{L_{5}}\widetilde{A_{5}}=\left[
\begin{array}{cccc}
1&2&3&4\\
4&0&3&2\\
0&3&2&4\\
6&4&5&5\\
\end{array}\right]
\end{align*}
are all NSC quasi-orthogonal matrix over $\mathbb{F}_{7}$. Similarly to the above proposition, we have the following proposition.

\begin{proposition}\label{proposition3.6}
Let $C_{i}$ be an $[n,t_{i},d_{i}]_{7}$ linear code with $C_{i}^{\perp}\subseteq C_{i}$ for each $i=1,2,3,4$. Then,

(1) The matrix-product code $C(\widetilde{L_{3}}\widetilde{A_{3}})=[C_{1},C_{2}] \widetilde{L_{3}}\widetilde{A_{3}}$ is an
$[2n,t_{1}+t_{2},\geq d]_{7}$ Euclidean dual-containing code, where $d=\min \{2d_{1},d_{2}\}$. Further, $C(\widetilde{L_{3}}\widetilde{A_{3}})$
generates an $[[2n,2(t_{1}+t_{2}-n),\geq d]]_{7}$ quantum code;

(2) The matrix-product code $C(\widetilde{L_{4}}\widetilde{A_{4}})=[C_{1},C_{2},C_{3}] \widetilde{L_{4}}\widetilde{A_{4}}$ is an
$[3n,t_{1}+t_{2}+t_{3},\geq d]_{7}$ Euclidean dual-containing code, where $d=\min \{3d_{1},2d_{2},d_{3}\}$. Further, $C(\widetilde{L_{4}}\widetilde{A_{4}})$ generates an $[[3n,2(t_{1}+t_{2}+t_{3})-3n,\geq d]]_{7}$ quantum code;

(3) The matrix-product code $C(\widetilde{L_{5}}\widetilde{A_{5}})=[C_{1},C_{2},C_{3},C_{4}] \widetilde{L_{5}}\widetilde{A_{5}}$ is an
$[4n,t_{1}+t_{2}+t_{3}+t_{4},\geq d]_{7}$ Euclidean dual-containing code, where $d=\min \{4d_{1},3d_{2},2d_{3},d_{4}\}$.
Further, $C(\widetilde{L_{5}}\widetilde{A_{5}})$ generates an $[[4n,2(t_{1}+t_{2}+t_{3}+t_{4}-2n),\geq d]]_{7}$ quantum code.
\end{proposition}

Now we are able to construct the $2\times 2$, $3\times 3$ and $4\times 4$ NSC quasi-orthogonal matrices over $\mathbb{F}_{9}$.
Suppose $\xi$ is a primitive element of $\mathbb{F}_{9}$. Denote 
\begin{equation*}
\widetilde{A_{6}}=\left[
\begin{array}{cc}
\xi^{2}&\xi^{2}\\
1&\xi^{2}\\
\end{array}\right],
\widetilde{A_{7}}=\left[
\begin{array}{ccc}
1&\xi^{2}&1\\
0&1&\xi\\
1&\xi&\xi^{2}\\
\end{array}\right],
\widetilde{A_{8}}=\left[
\begin{array}{cccc}
1&1&\xi^{2}&1\\
0&1&1&\xi^{2}\\
1&0&-1&\xi^{2}\\
0&0&0&1\\
\end{array}\right].
\end{equation*}
One can verify that $\widetilde{A_{6}}$, $\widetilde{A_{7}}$ and $\widetilde{A_{8}}$ are all NSC, and
all leading principal minors of
$\widetilde{A_{6}}\widetilde{A_{6}}^{T}$,
$\widetilde{A_{7}}\widetilde{A_{7}}^{T}$ and $\widetilde{A_{8}}\widetilde{A_{8}}^{T}$ are nonzero.
By Theorem \ref{theorem3.6}, we obtain lower unitriangular matrices
\begin{equation*}
\widetilde{L_{6}}=\left[
\begin{array}{cc}
1&0\\
1-\xi^{2}&1\\
\end{array}\right],
\widetilde{L_{7}}=\left[
\begin{array}{ccc}
1&0&0\\
-\xi^{2}-\xi&1&0\\
\xi^{3}+1&-\xi^{2}-\xi&1\\
\end{array}\right],
\widetilde{L_{8}}=\left[
\begin{array}{cccc}
1&0&0&0\\
1-\xi^{2}&1&0&0\\
-\xi^{2}-1&1&1&0\\
0&-\xi^{2}-1&1&1\\
\end{array}\right]
\end{equation*}
such that $\widetilde{L_{6}}\widetilde{A_{6}}\widetilde{A_{6}}^{T}\widetilde{L_{6}}^{T}=\mathrm {diag}(1,-\xi^{2})$, $\widetilde{L_{7}}\widetilde{A_{7}}\widetilde{A_{7}}^{T}\widetilde{L_{7}}^{T}=\mathrm {diag}(1,\xi^{3}-1,\xi^{3}-1)$,
$\widetilde{L_{8}}\widetilde{A_{8}}\widetilde{A_{8}}^{T}\widetilde{L_{8}}^{T}=\mathrm {diag}(-1,\xi^{2}+1,1-\xi^{2},-1)$.
Hence,
\begin{align*}
\widetilde{L_{6}}\widetilde{A_{6}}=\left[
\begin{array}{cc}
\xi^{2}&\xi^{2}\\
\xi^{2}-1&1-\xi^{2}\\
\end{array}\right],
\widetilde{L_{7}}\widetilde{A_{7}}=\left[
\begin{array}{ccc}
1&\xi^{2}&1\\
-\xi^{2}-\xi&-\xi^{3}-1&-\xi^{2}\\
\xi^{3}-1&-\xi&1\\
\end{array}\right],
\end{align*}
\begin{align*}
\widetilde{L_{8}}\widetilde{A_{8}}=\left[
\begin{array}{cccc}
1&1&\xi^{2}&1\\
1-\xi^{2}&-1-\xi^{2}&\xi^{2}-1&1\\
-\xi^{2}&-\xi^{2}&1-\xi^{2}&\xi^{2}-1\\
1&-\xi^{2}-1&1-\xi^{2}&-1\\
\end{array}\right]
\end{align*}
are all NSC quasi-orthogonal matrix over $\mathbb{F}_{9}$. This immediately derives the following proposition.

\begin{proposition}\label{proposition3.7}
Let $C_{i}$ be an $[n,t_{i},d_{i}]_{9}$ linear code with $C_{i}^{\perp}\subseteq C_{i}$ for each $i=1,2,3,4$. Then,

(1) The matrix-product code $C(\widetilde{L_{6}}\widetilde{A_{6}})=[C_{1},C_{2}] \widetilde{L_{6}}\widetilde{A_{6}}$ is an
$[2n,t_{1}+t_{2},\geq d]_{9}$ Euclidean dual-containing code, where $d=\min \{2d_{1},d_{2}\}$. Further, $C(\widetilde{L_{6}}\widetilde{A_{6}})$
generates an $[[2n,2(t_{1}+t_{2}-n),\geq d]]_{9}$ quantum code;

(2) The matrix-product code $C(\widetilde{L_{7}}\widetilde{A_{7}})=[C_{1},C_{2},C_{3}] \widetilde{L_{7}}\widetilde{A_{7}}$ is an
$[3n,t_{1}+t_{2}+t_{3},\geq d]_{9}$ Euclidean dual-containing code, where $d=\min \{3d_{1},2d_{2},d_{3}\}$. Further, $C(\widetilde{L_{7}}\widetilde{A_{7}})$
generates an $[[3n,2(t_{1}+t_{2}+t_{3})-3n,\geq d]]_{9}$ quantum code;

(3) The matrix-product code $C(\widetilde{L_{8}}\widetilde{A_{8}})=[C_{1},C_{2},C_{3},C_{4}] \widetilde{L_{8}}\widetilde{A_{8}}$ is an
$[4n,t_{1}+t_{2}+t_{3}+t_{4},\geq d]_{9}$ Euclidean dual-containing code, where $d=\min \{4d_{1},3d_{2},2d_{3},d_{4}\}$. Further,
$C(\widetilde{L_{8}}\widetilde{A_{8}})$ generates an $[[4n,2(t_{1}+t_{2}+t_{3}+t_{4}-2n),\geq d]]_{9}$ quantum code.
\end{proposition}

\section{Quantum codes from Hermitian dual-containing matrix-product codes}\label{section4}

\subsection{Basic concepts and properties}\label{subsection4.1}
Assume that $q$ is a prime power. Let $\mathbb{F}_{q^{2}}$ be the finite field with $q^{2}$ elements and let $\mathbb{F}_{q^{2}}^{\ast}$ denote the set of non-zero elements over $\mathbb{F}_{q^{2}}$. Let $\mathcal{M}(\mathbb{F}_{q^{2}},s\times t)$ be the set of $s\times t$ matrices over $\mathbb{F}_{q^{2}}$.
For any $a\in\mathbb{F}_{q^{2}}$, we define $\overline{a}=a^{q}$ the conjugate of $a$.
For any matrix $A=(a_{ij})\in \mathcal{M}(\mathbb{F}_{q^{2}},s\times t)$,
we define $A^{\dagger}=(\overline{a_{ji}})$ its conjugate transpose.

\begin{definition}\label{definition4.1}
For two vectors
$\mathbf{x}=(x_{1},x_{2},\ldots,x_{n})$, $\mathbf{y}=(y_{1},y_{2},\ldots,y_{n})\in \mathbb{F}_{q^{2}}^{n}$,
define
\begin{equation*}
(\mathbf{x},\mathbf{y})_{H}=\sum_{i=1}^n x_i \overline{y_{i}}.
\end{equation*}
as the \emph{Hermitian inner product} of $\mathbf{x}$ and $\mathbf{y}$.
For a linear code $C$ of length $n$ over $\mathbb{F}_{q^{2}}$, define its \emph{Hermitian dual code} as
\begin{equation*}
C^{\perp_{H}}=\{\mathbf{x}\in\mathbb{F}_{q^{2}}^{n}|(\mathbf{x},\mathbf{y})_{H}=0   \ \mathrm{for} \ \mathrm{all} \  \mathbf{y}\in C\}.
\end{equation*}
\end{definition}

\begin{definition}\label{definition4.2}
Let $C$ be a linear code over $\mathbb{F}_{q^{2}}^{n}$. Then,

(1) If $C\subseteq C^{\perp_{H}}$, then $C$ is called a \emph{Hermitian self-orthogonal code};

(2) If $C^{\perp_{H}}= C$, then $C$ is called a \emph{Hermitian self-dual code};

(3) If $C^{\perp_{H}}\subseteq C$, then $C$ is called a \emph{Hermitian dual-containing code}.
\end{definition}

For any vector $\mathbf{v}=(v_{1},v_{2},\ldots,v_{n})\in \mathbb{F}_{q^{2}}^{n}$, we define ${\mathbf{v}}^{q}=(v_{1}^{q},v_{2}^{q},\ldots,v_{n}^{q})$.
Let $S$ be a subset of $\mathbb{F}_{q^{2}}^{n}$ and define $S^{q}=\{{\mathbf{v}}^{q}|\mathbf{v}\in S\}$. Then, one can verify that $C^{\perp_{H}}=(C^{q})^{\perp}$ holds for any linear code $C$ over $\mathbb{F}_{q^{2}}$.
Therefore, we know that $C$ is Hermitian dual-containing if and only if
$(C^{q})^{\perp}\subseteq C$ if and only if $C^{\perp}\subseteq C^{q}$.

\begin{lemma}\label{lemma4.1}
(\!\!\cite{Ashikhmin2001}, Hermitian construction) If $C$ is an $[n,k,d]_{q^{2}}$ linear code with $C^{\perp_{H}}\subseteq C$, then there exists an
$[[n,2k-n,\geq d]]_{q}$ quantum code.
\end{lemma}

For a matrix-product code over $\mathbb{F}_{q^{2}}$, its Hermitian dual code has the following form.

\begin{lemma}\label{lemma4.2}
(\!\!\cite{Zhang2015qc}) Let $C_{i}$ be an $[n,t_{i},d_{i}]_{q^{2}}$ linear code, where $i=1,2,\ldots,k$. If $A\in
\mathcal{M}(\mathbb{F}_{q^{2}},k\times k)$ is non-singular, then
\begin{equation*}
([C_{1},C_{2},\ldots,C_{k}] A)^{\perp_{H}}=[C_{1}^{\perp_{H}},C_{2}^{\perp_{H}},\ldots,C_{k}^{\perp _{H}}](A^{-1})^\dagger.
\end{equation*}
\end{lemma}

\subsection{General approach for constructing quantum codes via Hermitian dual-containing matrix-product codes}\label{subsection4.2}

Let $$\tau=\left(
\begin{array}{cccc}
1& 2&  \cdots &k\\
i_{1}& i_{2}& \cdots &i_{k}  \\
\end{array}\right)$$
denote a permutation on $ \{1,2,\ldots,k\}$. Let $P_{\tau}$
be the $k\times k$ permutation matrix in which the $i_{j}$-th row
of the identity matrix $I_{k}$ is replaced with the $j$-th row of it for $j=1,2,\ldots,k$.

\begin{definition}\label{definition4.3}
Let $B\in \mathcal{M}(\mathbb{F}_{q^{2}},k\times k)$. If $B=DP_{\tau}$, where $D=\mathrm{diag}(d_{11},\ldots,d_{kk})$ with $d_{ii} \in \mathbb{F}_{q^{2}}^{\ast}$ for each $i$, then we call $B$ a \emph{monomial matrix} with respect to the permutation $\tau$.
\end{definition}

\begin{definition}\label{definition4.4}
Let $B\in \mathcal{M}(\mathbb{F}_{q^{2}},k\times k)$. If $BB^{\dag}$ is diagonal over $\mathbb{F}_{q}^{\ast}$,
then we call $B$ a \emph{quasi-unitary matrix} over $\mathbb{F}_{q^{2}}$.
If $BB^{\dag}=I_{k}$, then we call $B$ an \emph{unitary matrix} over $\mathbb{F}_{q^{2}}$.
\end{definition}

The following theorem gives a general approach for constructing $q$-ary quantum codes via Hermitian dual-containing matrix-product
codes over $\mathbb{F}_{q^{2}}$.

\begin{theorem}\label{theorem4.1}
(\!\!\cite{Cao2020Constructionof}) Let $C_{j}$ be an $[n,t_{j},d_{j}]_{q^{2}}$ linear code with
$C_{i_{j}}^{\perp_{H}}\subseteq C_{j}$ for $j=1,2,\ldots,k$. For any non-singular matrix $A\in\mathcal{M}(\mathbb{F}_{q^{2}},k\times k)$,
if $AA^{\dagger}$ is a monomial matrix with respect to the permutation $\tau$, then the matrix-product code
\begin{equation*}
C(A)=[C_{1},C_{2},\ldots,C_{k}]A
\end{equation*}
is an $[kn,\sum_{i=1}^k t_i,\geq d]_{q^{2}}$ Hermitian dual-containing code, where
$d=\min \limits_{1\leq i\leq k}\{D_{i}(A)d_{i}\}$. Further, $C(A)$ generates an $[[kn,2\sum_{i=1}^k t_i-kn,\geq d]]_{q}$ quantum code.
\end{theorem}

By Theorem \ref{theorem4.1}, we can construct a Hermitian dual-containing matrix-product code $C(A)$ over $\mathbb{F}_{q^{2}}$,
as long as the following two conditions hold:

(i) $AA^{\dag}$ is a monomial matrix with respect to $\tau$, where $\tau$ maps each $j\in \{1,2,\ldots,k\}$ to $i_{j}\in \{1,2,\ldots,k\}$;

(ii) The constituent codes $C_{1},C_{2},\ldots,C_{k}$ of $C(A)$ satisfy $C_{i_{j}}^{\perp_{H}}\subseteq C_{j}$ for each $j=1,2,\ldots,k$.

Further, we can obtain $q$-ary quantum codes by Hermitian construction. In fact, for a given permutation $\tau$, it is not difficult to find proper constituent codes satisfying (ii). The main difficulty lies in condition (i).
As far as we know, a general method for finding the matrix $A$ in condition (i) is still lacking at present.
In the remaining subsections of this section, we will study the constructions of quantum codes
when the defining matrix $A$ of $C(A)$ in turn satisfies different cases: (1) $A$ is a quasi-unitary matrix;
(2) $A$ is a NSC quasi-unitary matrix.

\subsection{Quantum codes related to quasi-unitary matrices}\label{subsection4.3}

When $\tau$ is an identity permutation, i.e., $\tau=(1)$, it follows from Theorem \ref{theorem4.1} that

\begin{corollary}\label{corollary4.1}
Let $C_{i}$ be an $[n,t_{i},d_{i}]_{q}$ linear code with $C_{i}^{\perp_{H}}\subseteq C_{i}$ for $i=1,2,\ldots,k$.
For any non-singular $A\in \mathcal{M}(\mathbb{F}_{q^{2}},k\times k)$,
if $A$ is quasi-unitary, then the matrix-product code
\begin{equation*}
C(A)=[C_{1},C_{2},\ldots,C_{k}]A
\end{equation*}
is an $[kn,\sum_{i=1}^k t_i,\geq d]_{q^{2}}$ Hermitian dual-containing code, where $d=\min \limits_{1\leq i\leq k}\{d_{i}D_{i}(A)\}$.
Further, $C(A)$ generates an $[[kn,2\sum_{i=1}^k t_i-kn,\geq d]]_{q}$ quantum code.
\end{corollary}

\subsubsection{Quantum codes related to quasi-unitary matrices for odd prime power $q$}\label{subsubsection4.3.1}

\begin{definition}\label{definition4.5}
(\!\!\cite{Giuzzi2000Hermitian}) Let $K$ be a field with character $\mathrm{char}(K)\neq 2$. Let $\sigma$ be a automorphism of $K$. If $\sigma\neq \mathrm{id}$ and $\sigma^{2}=\mathrm{id}$, then $\sigma$ is called an \emph{involution}.
\end{definition}

\begin{definition}\label{definition4.6}
(\!\!\cite{Giuzzi2000Hermitian}) For any matrix $A=(a_{ij})$, define $A^{\sigma}=(\sigma(a_{ij}))$. Given a matrix $H\in\mathcal{M}(K,k\times k)$. If $H^{T}=H^{\sigma}$,
then $H$ is called a \emph{Hermitian matrix} with respect to the automorphism $\sigma$.
\end{definition}

\begin{lemma}\label{lemma4.3}
(\!\!\cite{Giuzzi2000Hermitian}) Let $K$ be a field with character $\mathrm{char}(K)\neq 2$. Let $\sigma$ be an involution of $K$ and let $K_{0}$ be the fixed subfield of
$\sigma$ in $K$. If $H\in\mathcal{M}(K,k\times k)$ is a Hermitian matrix with respect to $\sigma$
and $\mathrm{rank}(H)=k-t$, then there exists a non-singular matrix
$M\in\mathcal{M}(K,k\times k)$ such that
\begin{equation*}
M^{\sigma}HM^{T}=\mathrm{diag}(h_{1},\ldots,h_{k-t},0,\ldots,0),
\end{equation*}
where $h_{i}\in K_{0}$ for $i=1,2,\ldots,k-t$.
\end{lemma}

Based on the above facts, we will prove that there exist quasi-unitary matrices and unitary matrices over $\mathbb{F}_{q^{2}}$
when $q$ is an odd prime power.

\begin{proposition}\label{proposition4.1}
Let $q$ be an odd prime power. Then, for any non-singular matrix $A\in\mathcal{M}(\mathbb{F}_{q^{2}},k\times k)$, there exists a non-singular matrix $N\in\mathcal{M}(\mathbb{F}_{q^{2}},k\times k)$
such that $NAA^{\dag}N^{\dag}$ is diagonal over $\mathbb{F}_{q}^{\ast}$, i.e., $NA$ is quasi-unitary over $\mathbb{F}_{q^{2}}$.
Moreover, there exists a non-singular matrix $N'\in\mathcal{M}(\mathbb{F}_{q^{2}},k\times k)$ such that $N'A$ is unitary over $\mathbb{F}_{q^{2}}$.
\end{proposition}

\begin{proof}
In Definition \ref{definition4.5}, let $K=\mathbb{F}_{q^{2}}$ and $\sigma=\sigma_{1}:x\mapsto x^{q}$. One can verify that $\sigma_{1}$
is an involution.
Hence, $H$ is a Hermitian matrix with respect to $\sigma_{1}$ if and only if $H^{\dag}=H$.
Since $A^{\dag}=(A^{\sigma_{1}})^{T}$, we know $(AA^{\dag})^{T}=A^{\sigma_{1}}A^{T}=(AA^{\dag})^{\sigma_{1}}$.
Then, $AA^{\dag}$ is a Hermitian matrix with respect to $\sigma_{1}$.
As $AA^{\dag}$ is non-singular, it follows from Lemma \ref{lemma4.3} that there exists a non-singular matrix $N$
such that $NAA^{\dag}N^{\dag}$ is diagonal over $\mathbb{F}_{q^{2}}^{\ast}$.
Write $NAA^{\dag}N^{\dag}=\mbox{diag}(r_{1},r_{2},\ldots,r_{k})$, where each $r_{i}\in\mathbb{F}_{q^{2}}^{\ast}$.
By $NAA^{\dag}N^{\dag}=(NAA^{\dag}N^{\dag})^{\dag}$ we know that $r_{i}=r_{i}^{q}$, then $r_{i}\in\mathbb{F}_{q}^{\ast}$.
Thus, $NAA^{\dag}N^{\dag}$ is diagonal over $\mathbb{F}_{q}^{\ast}$, i.e., $NA$ is quasi-unitary over $\mathbb{F}_{q^{2}}$.

Moreover, by $r_{i}\in\mathbb{F}_{q}^{\ast}$ we know that there exists $s_{i}\in\mathbb{F}_{q^{2}}^{\ast}$ such that $r_{i}=s_{i}^{q+1}$.
Let $N'=D_{1}N$, where $D_{1}=\mbox{diag}(s_{1}^{-1},s_{2}^{-1},\ldots,s_{k}^{-1})$. Then,
$N'AA^{\dag}(N')^{\dag}=I_{k}$, i.e., $N'A$ is unitary over $\mathbb{F}_{q^{2}}$. This completes the proof. $\hfill\square$
\end{proof}

By Proposition \ref{proposition4.1}, we can construct Hermitian dual-containing matrix-product codes and obtain the corresponding
quantum codes in the following theorem.

\begin{theorem}\label{theorem4.2}
Let $q$ be an odd prime power. Let $C_{i}$ be an $[n,t_{i},d_{i}]_{q^{2}}$ linear code with $C_{i}^{\bot_{H}}\subseteq C_{i}$ for $i=1,2,\ldots,k$.
Then, for any non-singular matrix $A\in\mathcal{M}(\mathbb{F}_{q^{2}},k\times k)$, there exist non-singular matrices $N, N'\in\mathcal{M}(\mathbb{F}_{q^{2}},k\times k)$
such that the matrix-product codes
\begin{equation*}
C(NA)=[C_{1},C_{2},\ldots,C_{k}]NA
\end{equation*}
and
\begin{equation*}
C(N'A)=[C_{1},C_{2},\ldots,C_{k}]N'A
\end{equation*}
are $[kn,\sum_{i=1}^{k} t_i,\geq d']_{q^{2}}$ Hermitian dual-containing code and $[kn,\sum_{i=1}^{k} t_i,\geq d'']_{q^{2}}$
Hermitian dual-containing code, respectively, where $d'=\min \limits_{1\leq i\leq k}\{d_{i}D_{i}(NA)\}$ and
$d''=\min \limits_{1\leq i\leq k}\{d_{i}D_{i}(N'A)\}$. Further, $C(NA)$ and $C(N'A)$ can generate quantum codes with parameters
$[[kn,2\sum_{i=1}^{k} t_i-kn,\geq d']]_{q}$ and $[[kn,2\sum_{i=1}^{k} t_i-kn,\geq d'']]_{q}$, respectively.
\end{theorem}

\begin{proof}
It is immediately obtained by Corollary \ref{corollary4.1} and Proposition \ref{proposition4.1}. $\hfill\square$
\end{proof}

\subsubsection{Quantum codes related to a $2^{m}\times 2^{m}$ quasi-unitary matrices}\label{subsubsection4.3.2}

\begin{proposition}\label{proposition4.2}
In the finite field $\mathbb{F}_{q^{2}}$, suppose $c=c_{1}^{q+1}+\cdots+c_{2^{m}}^{q+1}\neq 0$.
Then, there exists a matrix $S=(s_{i,j})\in \mathcal{M}(\mathbb{F}_{q^{2}},2^{m}\times 2^{m})$ such that $S^{\dag}S=SS^{\dag}=cI_{2^{m}}$,
i.e., $S$ is quasi-unitary over $\mathbb{F}_{q^{2}}$, where $(c_{1},\ldots,c_{2^{m}})$ denotes the first row of $S$.
\end{proposition}

\begin{proof}
Assume that the result holds for $m-1$. Write $c=a+b$, where
\begin{equation*}
a=c_{1}^{q+1}+\cdots+c_{2^{m-1}}^{q+1},
\end{equation*}
\begin{equation*}
b=c_{2^{m-1}+1}^{q+1}+\cdots+c_{2^{m}}^{q+1}.
\end{equation*}
Then, there exist $A,B\in \mathcal{M}(\mathbb{F}_{q^{2}},2^{m-1}\times 2^{m-1})$ such that
\begin{equation*}
AA^{\dag}=A^{\dag}A=aI_{2^{m-1}},
\end{equation*}
\begin{equation*}
BB^{\dag}=B^{\dag}B=bI_{2^{m-1}},
\end{equation*}
where $(c_{1},\ldots,c_{2^{m-1}})$ and $(c_{2^{m-1}+1},\ldots,c_{2^{m}})$ are the first row of $A$ and $B$, respectively.

Now let us discuss the following two cases.

{\bfseries Case (1):} When $a\neq 0$, take
\begin{equation*}
S=\left[
\begin{array}{cc}
A&  B\\
-a^{-1}A^{\dag}B^{\dag}A&A^{\dag}  \\
\end{array}\right].
\end{equation*}
Then, we have $S^{\dag}{S}=SS^{\dag}=cI_{2^{m}}$, and $(c_{1},\ldots,c_{2^{m}})$ is the first row of $S$.

{\bfseries Case (2):} When $b\neq 0$, take
\begin{equation*}
S=\left[
\begin{array}{cc}
B&  A\\
-b^{-1}B^{\dag}A^{\dag}B&B^{\dag}  \\
\end{array}\right].
\end{equation*}
Then, we have $S^{\dag}{S}=SS^{\dag}=cI_{2^{m}}$, and $(c_{1},\ldots,c_{2^{m}})$ is the first row of $S$.

Therefore, $S$ is quasi-unitary over $\mathbb{F}_{q^{2}}$. This completes the proof. $\hfill\square$
\end{proof}

By Corollary \ref{corollary4.1} and Proposition \ref{proposition4.2}, we can directly obtain the following theorem.

\begin{theorem}\label{theorem4.3}
Let $C_{i}$ be an $[n,t_{i},d_{i}]_{q}$ linear code with $C_{i}^{\bot_{H}}\subseteq C_{i}$ for $i=1,2,\ldots,2^{m}$.
In $\mathbb{F}_{q^{2}}$, write $c=c_{1}^{q+1}+\cdots+c_{2^{m}}^{q+1}\neq 0$. Then, there exists a matrix
$S\in \mathcal{M}(\mathbb{F}_{q^{2}},2^{m}\times 2^{m})$ such that the matrix-product code
\begin{equation*}
C(S)=[C_{1},C_{2},\ldots,C_{2^{m}}]S
\end{equation*}
is an $[2^{m}n,\sum_{i=1}^{2^{m}} t_i,\geq d]_{q^{2}}$ Hermitian dual-containing code,
where $d=\min \limits_{1\leq i\leq 2^{m}}\{d_{i}D_{i}(S)\}$. Further, $C(S)$ generates an
$[[2^{m}n,2\sum_{i=1}^{2^{m}} t_i-2^{m}n,\geq d]]_{q}$ quantum code.
\end{theorem}

\subsection{Quantum codes related to NSC quasi-unitary matrices}\label{subsection4.4}

Recall that in Corollary \ref{corollary4.1} when the defining matrix $A$ is quasi-unitary, we can construct a
Hermitian dual-containing matrix-product code.
Further, it can generate an $[[kn,2\sum_{i=1}^k t_i-kn,\geq d]]_{q}$ quantum code from Hermitian construction, where
$d=\min \limits_{1\leq i\leq k}\{d_{i}D_{i}(A)\}$. Here $D_{i}(A)\leq k+1-i$.
Similar to Subsection \ref{subsection3.4}, we wish to make the minimum distance lower bound of the constructed matrix-product codes
as large as possible.
In other words, we need to find the quasi-unitary matrix $A$ satisfying $D_{i}(A)=k+1-i$.
Similar to Subsection \ref{subsection3.4}, we can consider NSC matrices.
For a matrix $A$, if it is both NSC and quasi-unitary, then
we call $A$ \emph{NSC quasi-unitary}.

The following theorem gives a constructive method for constructing general quasi-unitary matrices and NSC quasi-unitary matrices.

\begin{theorem}\label{theorem4.4}
(\!\!\cite{Cao2020Constructioncaowang}) Let $A\in\mathcal{M}(\mathbb{F}_{q^{2}},k\times k)$ be non-singular.
If all leading principal minors of $AA^{\dag}$ are nonzero,
then there exists a lower unitriangular matrix $L$ such that $LA$ is quasi-unitary over $\mathbb{F}_{q^{2}}$.
Further, if $A$ is NSC, then $LA$ is NSC quasi-unitary.
\end{theorem}

Next, let $\alpha$ be a primitive element of $\mathbb{F}_{q^{2}}$. Assume that $k\mid (q+1)$ and write $\beta_{i}=\alpha^{\frac{q^{2}-1}{k}i}$
for $i=0,1,\ldots,k-1$.
If we take $L=I_{k}$ in Theorem \ref{theorem4.4}, then we get the following corollary.

\begin{corollary}\label{corollary4.3}
(\!\!\cite{Jitman2017}) Let $k|(q+1)$ and $M=(\beta_{j-1}^{i-1})_{i,j=1}^{k}$. Then, we have $MM^{\dag}=kI_{k}$, i.e., $M$ is NSC quasi-unitary.
\end{corollary}

According to Theorem \ref{theorem4.4}, we have the following theorem.

\begin{theorem}\label{theorem4.5}
(\!\!\cite{Cao2020Constructioncaowang}) Let $C_{i}$ be an $[n,t_{i},d_{i}]_{q^{2}}$ linear code with $C_{i}^{\perp_{H}}\subseteq C_{i}$ for each $i=1,2,\ldots,k$.
For any NSC matrix $A\in\mathcal{M}(\mathbb{F}_{q^{2}},k\times k)$, if all leading principal minors of $AA^{\dag}$ are nonzero,
then there exists a lower unitriangular matrix $L$ such that the matrix-product code
\begin{equation*}
C(LA)=[C_{1},C_{2},\ldots,C_{k}]LA
\end{equation*}
is an $[kn,\sum_{i=1}^k t_i,\geq d]_{q^{2}}$ Hermitian dual-containing code,
where $d=\min \limits_{1\leq i\leq k}\{(k+1-i)d_{i}\}$.
Further, $C(LA)$ generates an $[[kn,2\sum_{i=1}^k t_i-kn,\geq d]]_{q}$ quantum code.
\end{theorem}

\begin{corollary}\label{corollary4.5}
Let $k|(q+1)$ and $M$ be defined as in Corollary \ref{corollary4.3}. Let $C_{i}$ be an $[n,t_{i},d_{i}]_{q^{2}}$ linear code with
$C_{i}^{\perp_{H}}\subseteq C_{i}$ for each $i=1,2,\ldots,k$.
Then, the matrix-product code
\begin{equation*}
C(M)=[C_{1},C_{2},\ldots,C_{k}]M
\end{equation*}
is an $[kn,\sum_{i=1}^k t_i,\geq d]_{q^{2}}$ Hermitian dual-containing code,
where $d=\min \limits_{1\leq i\leq k}\{(k+1-i)d_{i}\}$.
Further, $C(M)$ generates an $[[kn,2\sum_{i=1}^k t_i-kn,\geq d]]_{q}$ quantum code by Hermitian construction.
\end{corollary}

From Theorem \ref{theorem4.5}, we have the following corollary.

\begin{corollary}\label{corollary4.6}
Let $C_{i}$ be an $[n,t_{i},d_{i}]_{q^{2}}$ linear code with $C_{i}^{\perp_{H}}\subseteq C_{i}$ for $i=1,2,3,4$. Then,

(1) There exists an $2\times 2$ NSC quasi-unitary matrix $\widehat{L_{1}}\widehat{A_{1}}$ over $\mathbb{F}_{q^{2}}$,
such that
\begin{equation*}
C(\widehat{L_{1}}\widehat{A_{1}})=[C_{1},C_{2}]\widehat{L_{1}}\widehat{A_{1}}
\end{equation*}
is an $[2n,t_{1}+t_{2},\geq d]_{q^{2}}$ Hermitian dual-containing matrix-product code, where $d=\min \{2d_{1},d_{2}\}$.
Further, $C(\widehat{L_{1}}\widehat{A_{1}})$ generates an $[[2n,2(t_{1}+t_{2}-n),\geq d]]_{q}$ quantum code.

(2) There exists an $3\times 3$ NSC quasi-unitary matrix $\widehat{L_{2}}\widehat{A_{2}}$ over $\mathbb{F}_{q^{2}}$,
such that
\begin{equation*}
C(\widehat{L_{2}}\widehat{A_{2}})=[C_{1},C_{2},C_{3}]\widehat{L_{2}}\widehat{A_{2}}
\end{equation*}
is an $[3n,t_{1}+t_{2}+t_{3},\geq d]_{q^{2}}$ Hermitian dual-containing matrix-product code, where $d=\min \{3d_{1},2d_{2},d_{3}\}$.
Further, $C(\widehat{L_{2}}\widehat{A_{2}})$ generates an $[[3n,2(t_{1}+t_{2}+t_{3})-3n,\geq d]]_{q}$ quantum code.

(3) When $q\geq5$, there exists an $4\times 4$ NSC quasi-unitary matrix $\widehat{L_{3}}\widehat{A_{3}}$ over $\mathbb{F}_{q^{2}}$,
such that
\begin{equation*}
C(\widehat{L_{3}}\widehat{A_{3}})=[C_{1},C_{2},C_{3},C_{4}]\widehat{L_{3}}\widehat{A_{3}}
\end{equation*}
is an $[4n,t_{1}+t_{2}+t_{3}+t_{4},\geq d]_{q^{2}}$ Hermitian dual-containing matrix-product code, where $d=\min \{4d_{1},3d_{2},2d_{3},d_{4}\}$.
Further, $C(\widehat{L_{3}}\widehat{A_{3}})$ generates an $[[4n,2(t_{1}+t_{2}+t_{3}+t_{4})-4n,\geq d]]_{q}$ quantum code.
\end{corollary}

\begin{proof}
(1) Define
\begin{equation*}
\mathcal{R}_{1}=\{x|x\in\mathbb{F}_{q^{2}}^{\ast},x^{2}\neq 1,x^{q+1}\neq -1\}.
\end{equation*}
Clearly, this set is nonempty. For any $a_{1}\in\mathcal{R}_{1}$, take
\begin{equation*}
\widehat{A_{1}}=\left[
\begin{array}{cc}
1&a_{1}\\
a_{1}&1\\
\end{array}\right],
\end{equation*}
then $\widehat{A_{1}}$ is NSC and all leading principal minors of $\widehat{A_{1}}\widehat{A_{1}}^{\dag}$ are nonzero.
By Theorem \ref{theorem4.4}, there exists a lower unitriangular matrix
\begin{equation*}
\widehat{L_{1}}=\left[
\begin{array}{cc}
1&0\\
-(a_{1}^{q}+a_{1})(1+a_{1}^{q+1})^{-1}&1\\
\end{array}\right]
\end{equation*}
such that
\begin{equation*}
\widehat{L_{1}}\widehat{A_{1}}\widehat{A_{1}}^{\dag}\widehat{L_{1}}^{\dag}=\mbox{diag}\big(1+a_{1}^{q+1},(1-a_{1}^{2})(1-a_{1}^{2q})(1+a_{1}^{q+1})^{-1} \big).
\end{equation*}
Hence, $\widehat{L_{1}}\widehat{A_{1}}$ is an $2\times 2$ NSC quasi-unitary matrix over $\mathbb{F}_{q^{2}}$.
Using Theorem \ref{theorem4.5}, we know that $C(\widehat{L_{1}}\widehat{A_{1}})$ is an $[2n,t_{1}+t_{2},\geq d]_{q^{2}}$
Hermitian dual-containing code. Further, it generates an $[[2n,2(t_{1}+t_{2}-n),\geq d]]_{q}$ quantum code by Hermitian construction,
where $d=\min \{2d_{1},d_{2}\}$.

(2) Define
\begin{equation*}
\mathcal{R}_{2}=\{x|x\in\mathbb{F}_{q^{2}}^{\ast}\backslash\{1\},x^{q+1}+2\neq 0,x^{q+1}-x^{q}-x+3\neq 0\}.
\end{equation*}
Clearly, this set is nonempty. For any $a_{2}\in\mathcal{R}_{2}$, take
\begin{equation*}
\widehat{A_{2}}=\left[
\begin{array}{ccc}
1&1&a_{2}\\
1&0&1\\
0&0&1\\
\end{array}\right],
\end{equation*}
then $\widehat{A_{2}}$ is NSC and all leading principal minors of $\widehat{A_{2}}\widehat{A_{2}}^{\dag}$ are nonzero.
By Theorem \ref{theorem4.4}, there exists a lower unitriangular matrix
\begin{equation*}
\widehat{L_{2}}=\left[
\begin{array}{ccc}
1&0&0\\
-(a_{2}^{q}+1)(a_{2}^{q+1}+2)^{-1}&1&0\\
(1-a_{2}^{q})p_{1}&(a_{2}^{q}-2)p_1&1\\
\end{array}\right]
\end{equation*}
such that
\begin{equation*}
\widehat{L_{2}}\widehat{A_{2}}\widehat{A_{2}}^{\dag}\widehat{L_{2}}^{\dag}=\mbox{diag}\big(a_{2}^{q+1}+2,(a_{2}^{q+1}+2)^{-1}p_{1}^{-1},p_{1} \big),
\end{equation*}
where $p_{1}=(a_{2}^{q+1}-a_{2}^{q}-a_{2}+3)^{-1}$.
Hence, $\widehat{L_{2}}\widehat{A_{2}}$ is an $3\times 3$ NSC quasi-unitary matrix over $\mathbb{F}_{q^{2}}$.
Using Theorem \ref{theorem4.5}, we know that $C(\widehat{L_{2}}\widehat{A_{2}})$ is an $[3n,t_{1}+t_{2}+t_{3},\geq d]_{q^{2}}$
Hermitian dual-containing code. Further, it generates an $[[3n,2(t_{1}+t_{2}+t_{3})-3n,\geq d]]_{q}$ quantum code by Hermitian construction,
where $d=\min \{3d_{1},2d_{2},d_{3}\}$.

(3) When $q\geq5$, define
\begin{equation*}
\mathcal{R}_{3}=\{x|x\in\mathbb{F}_{q^{2}}^{\ast}\backslash\{\pm1,3\},x^{q+1}+3\neq 0,2x^{q+1}+9\neq 0,x^{q+1}-3x^{q}-3x+15\neq 0\}.
\end{equation*}
Clearly, this set is nonempty. For any $a_{3}\in\mathcal{R}_{3}$, take
\begin{equation*}
\widehat{A_{3}}=\left[
\begin{array}{cccc}
1&1&1&a_{3}\\
1&0&-1&1\\
0&0&1&1\\
0&0&0&1\\
\end{array}\right],
\end{equation*}
then $\widehat{A_{3}}$ is NSC and all leading principal minors of $\widehat{A_{3}}\widehat{A_{3}}^{\dag}$ are nonzero.
By Theorem \ref{theorem4.4}, there exists a lower unitriangular matrix
\begin{equation*}
\widehat{L_{3}}=\left[
\begin{array}{cccc}
1&0&0&0\\
-a_{3}^{q}p_{4}&1&0&0\\
-3p_{3}p_{5}&a_{3}p_{3}p_{5}&1&0\\
(3-a_{3}^{q})p_{2}&(a_{3}^{q}-5)p_{2}&(2a_{3}^{q}-9)p_{2}&1\\
\end{array}\right]
\end{equation*}
such that
\begin{equation*}
\widehat{L_{3}}\widehat{A_{3}}\widehat{A_{3}}^{\dag}\widehat{L_{3}}^{\dag}=\mbox {diag}(p_{4}^{-1},p_{4}p_{3}^{-1},p_{3}p_{2}^{-1},p_{2}),
\end{equation*}
where $p_{2}=(a_{3}^{q+1}-3a_{3}^{q}-3a_{3}+15)^{-1}$, $p_{3}=(2a_{3}^{q+1}+9)^{-1}$, $p_{4}=(a_{3}^{q+1}+3)^{-1}$ and $p_{5}=a_{3}^{q}+1$.
Hence, when $q\geq5$, $\widehat{L_{3}}\widehat{A_{3}}$ is an $4\times 4$ NSC quasi-unitary matrix over $\mathbb{F}_{q^{2}}$.
Using Theorem \ref{theorem4.5}, we know that $C(\widehat{L_{3}}\widehat{A_{3}})$ is an $[4n,t_{1}+t_{2}+t_{3}+t_{4},\geq d]_{q^{2}}$
Hermitian dual-containing code. Further, it generates an $[[4n,2(t_{1}+t_{2}+t_{3}+t_{4})-4n,\geq d]]_{q}$ quantum code by Hermitian construction,
where $d=\min \{4d_{1},3d_{2},2d_{3},d_{4}\}$. $\hfill\square$
\end{proof}

\begin{remark}\label{remark4.3}
In Remark \ref{remark4.5}, we will give a different manner to construct the $3\times 3$ NSC quasi-unitary matrices over $\mathbb{F}_{q^{2}}$.
\end{remark}

\subsection{Quantum codes related to $k\times k$ NSC quasi-unitary matrices for any $k<q$}\label{subsection4.5}

For a nonzero polynomial $f(x_{1},\ldots,x_{k})\in \mathbb{F}_{q}[x_{1},\ldots,x_{k}]$,
we use $\mbox {deg}(f;x_{i})$ to denote the degree of $f(x_{1},\ldots,x_{k})$ in $x_{i}$.
First, we obtain the following proposition.

\begin{proposition}\label{proposition4.3}
(\cite{Cao2020Constructioncaowang}) For a nonzero polynomial $f(x_{1},\ldots,x_{k})\in \mathbb{F}_{q}[x_{1},\ldots,x_{k}]$, the following two statements are equivalent:

(1) For any $a_{1},\ldots,a_{k}\in \mathbb{F}_{q}$, $f(a_{1},\ldots,a_{k})=0$;

(2) There exist polynomials $g_{i}(x_{1},\ldots,x_{k}),i=1,\ldots,k$ in $\mathbb{F}_{q}[x_{1},\ldots,x_{k}]$ such that
$$f(x_{1},\ldots,x_{k})=\sum_{i=1}^k g_{i}(x_{1},\ldots,x_{k})\cdot(x_{i}^{q}-x_{i})$$
with $\mathrm {deg}(g_{j};x_{i})<q$ for each $i<j$.
\end{proposition}

By Proposition \ref{proposition4.3}, we obtain the following proposition.

\begin{proposition}\label{proposition4.4}
(\!\!\cite{Cao2020Constructioncaowang}) Let $f_{1}(x_{1},\ldots,x_{k}),f_{2}(x_{1},\ldots,x_{k}),\ldots,f_{v}(x_{1},\ldots,x_{k})$ be
$v$ nonzero polynomials in $\mathbb{F}_{q}[x_{1},\ldots,x_{k}]$. If
$\sum_{j=1}^v \mathrm {deg}(f_{j};x_{i})<q$ holds for each $i$, then there exist $a_{1},\ldots,a_{k}\in \mathbb{F}_{q}$ such that $f_{j}(a_{1},\ldots,a_{k})\neq 0$ holds for each $j$.
\end{proposition}

Denote by $A\dbinom{i_{1} \cdots i_{u}}{j_{1} \cdots j_{u}}$ the $u\times u$ matrix consisting of the $i_{1},\ldots,i_{u}$ rows and the
$j_{1},\ldots,j_{u}$ columns of $A$.
Obviously, $A\dbinom{1 \cdots u}{j_{1} \cdots j_{u}}$ is just the matrix
$A(j_{1},\ldots,j_{u})$ in Definition \ref{definition3.6}.

Now let us recall the Cauchy-Binet formula.

\begin{lemma}\label{lemma4.4}
(\!\!\cite{Horn2012Matrix}, Cauchy-Binet formula) Let $X=(x_{ij})$ be an $s\times n$ matrix and $Y=(y_{ij})$ be an $n\times s$ matrix.
Then, for any positive integer $u\leq s$,

(1) If $u>n$, then all $u\times u$ sub-determinant of $XY$ is $0$;

(2) If $u\leq n$, then the $u\times u$ sub-determinant consisting of the $i_{1},\ldots,i_{u}$ rows and the $j_{1},\ldots,j_{u}$ columns of $XY$
is
\begin{equation*}
\bigg |(XY)\dbinom{i_{1} \cdots i_{u}}{j_{1} \cdots j_{u}}\bigg|=\sum_{1\leq v_{1}<\cdots<v_{u}\leq n}\bigg |X\dbinom{i_{1} \cdots i_{u}}{v_{1} \cdots v_{u}}\bigg|\cdot \bigg |Y\dbinom{v_{1} \cdots v_{u}}{j_{1} \cdots j_{u}}\bigg|.
\end{equation*}
\end{lemma}

\vspace{6pt}
Using Proposition \ref{proposition4.4} and Lemma \ref{lemma4.4}, we can obtain the following proposition.

\begin{proposition}\label{proposition4.5}
(\!\!\cite{Cao2020Constructioncaowang}) Let $k$ be a positive integer with $k<q$. Then, for any non-singular matrix $A\in\mathcal{M}(\mathbb{F}_{q^{2}},k\times k)$, there exist
$\lambda_{1},\ldots,\lambda_{k}\in\mathbb{F}_{q^{2}}^{\ast}$ such that all leading principal minors of $BB^{\dag}$ are nonzero, where
$B=A\cdot\mathrm{diag}(\lambda_{1},\ldots,\lambda_{k})$.
\end{proposition}

By proposition \ref{proposition4.5}, we give the following proposition.

\begin{proposition}\label{proposition4.6}
(\!\!\cite{Cao2020Constructioncaowang}) Let $k$ be a positive integer with $k<q$. Then, for any NSC matrix $A\in\mathcal{M}(\mathbb{F}_{q^{2}},k\times k)$, there exist
$\lambda_{1},\ldots,\lambda_{k}\in\mathbb{F}_{q^{2}}^{\ast}$ such that the matrix
$B=A\cdot\mathrm{diag}(\lambda_{1},\ldots,\lambda_{k})$ satisfies the following properties:

(1) All leading principal minors of $BB^{\dag}$ are nonzero;

(2) $B\in\mathcal{M}(\mathbb{F}_{q^{2}},k\times k)$ is NSC.
\end{proposition}

Now, by using Proposition \ref{proposition4.6} and Theorem \ref{theorem4.4}, we obtain new classes of NSC quasi-unitary matrices as follows.

\begin{theorem}\label{theorem4.6}
(\!\!\cite{Cao2020Constructioncaowang}) There exist $k\times k$ NSC quasi-unitary matrices over $\mathbb{F}_{q^{2}}$ for all $k<q$.
\end{theorem}

\begin{remark}\label{remark4.4}
Up to now, the known (NSC) quasi-unitary matrices over $\mathbb{F}_{q^{2}}$ for constructing $q$-ary quantum codes via Hermitian
dual-containing matrix-product codes can be summarized as follows.
\begin{itemize}

\item Refs \cite{Jitman2017,Zhang2015qc,Liu2019Entanglement-assisted} constructed $2\times 2$
     NSC quasi-unitary matrices;

\item For any odd prime power $q$, Refs \cite{Liu2018On,Liu2019Entanglement-assisted}
      constructed $4\times 4$ quasi-unitary matrices that are not NSC;

\item Ref \cite{Jitman2017} constructed $k\times k$ NSC quasi-unitary matrices for $k|(q+1)$ (see also Corollary \ref{corollary4.3}),
     which can be deduced by the constructive method in Theorem \ref{theorem4.4};

\item In \cite{Cao2020Constructioncaowang}, the author of this article and his cooperators constructed $3\times 3$ NSC quasi-unitary matrices
     (see also Corollary \ref{corollary4.6}(2)). What's more, they constructed $k\times k$ NSC quasi-unitary matrices for any $k<q$
     (see also Theorem \ref{theorem4.6}).

\end{itemize}
Compared with the quasi-unitary matrices in \cite{Jitman2017,Zhang2015qc,Liu2019Entanglement-assisted,Liu2018On},
quasi-unitary matrices in \cite{Cao2020Constructioncaowang} are all NSC and their range of orders is much broader.
\end{remark}

\begin{remark}\label{remark4.5}
In Corollary \ref{corollary4.6}(2), we prove the existence of the $3\times 3$ NSC quasi-unitary matrices over $\mathbb{F}_{q^{2}}$
Here, we give a new manner to construct them. The detailed procedures are as follows.

{\bfseries Step (a):} By Corollary \ref{corollary4.3}, we know that there exist $3\times 3$ NSC quasi-unitary matrices over $\mathbb{F}_{2^{2}}$;

{\bfseries Step (b):} It follows from Theorem \ref{theorem4.6} that there exist $3\times 3$ NSC quasi-unitary matrices over $\mathbb{F}_{q^{2}}$
for any $q\geq 4$;

{\bfseries Step (c):} Define
\begin{equation*}
\mathcal{W}=\{x\in\mathbb{F}_{3^{2}}|x^{4}-1\neq 0,x^{4}-x^{3}-x\neq 0\}.
\end{equation*}
We know this set is nonempty. Take any $a\in \mathcal{W}$ and denote by
\begin{equation*}
A=\left[
\begin{array}{ccc}
1&1&a\\
1&0&1\\
0&0&1\\
\end{array}\right].
\end{equation*}
One can verify that $A$ is NSC and all leading principal minors of $AA^{\dag}$ are nonzero. By Theorem \ref{theorem4.4},
there exists a lower unitriangular matrix $L$ such that
$LA$ is NSC quasi-unitary. Hence, there exist $3\times 3$ NSC quasi-unitary matrices over $\mathbb{F}_{3^{2}}$.

In terms of Steps (a)-(c), we know that there always exist $3\times 3$ NSC quasi-unitary matrices in any finite field $\mathbb{F}_{q^{2}}$.
\end{remark}

Using Corollary \ref{corollary4.6}(2), Theorem \ref{theorem4.6} and Hermitian construction,
we obtain new classes of good quantum codes as follows.

\begin{theorem}\label{theorem4.7}
(\!\!\cite{Cao2020Constructioncaowang}) Let $C_{i}$ be an $[n,t_{i},d_{i}]_{q^{2}}$ linear code with $C_{i}^{\perp_{H}}\subseteq C_{i}$
for $i=1,\ldots,k$. Then,

(1) For $k=3$, there exists an $[[3n,2\sum_{i=1}^3 t_i-3n,\geq d]]_{q}$ quantum code, where
$d=\min \limits_{1\leq i\leq 3}\{(4-i)d_{i}\}$.

(2) For each positive integer $k$ with $k<q$, there exists an $[[kn,2\sum_{i=1}^k t_i-kn,\geq d]]_{q}$ quantum code, where
$d=\min \limits_{1\leq i\leq k}\{(k+1-i)d_{i}\}$.
\end{theorem}

\section{Concluding remarks}\label{section5}

It should be pointed out that the methods in Theorems \ref{theorem3.6} and \ref{theorem4.4}
and the NSC quasi-orthogonal (resp. NSC quasi-unitary) matrices constructed from them have the following potential:
\begin{itemize}
\item By these matrices, it is possible to obtain good entanglement-assisted quantum error-correcting codes
(EAQECCs) from matrix-product codes. For more information on EAQECCs, see \cite{Brun2006,Wilde2008,Wilde2010Entanglement-assisted,Wilde2014Entanglement-assisted,Guenda2018,Liu2020aa,Liu2020ab,Li2013,Hsieh2007General,
Cao2021MDS}.

\item By these matrices, it is possible to obtain good LCD codes from matrix-product codes.
\end{itemize}

For $q=2$, one can check that there are no $2n\times 2n$ NSC quasi-unitary matrices over $\mathbb{F}_{2^{2}}$ for any positive integer $n$.
By Corollary \ref{corollary4.3}, Corollary \ref{corollary4.6}(2) (or Remark \ref{remark4.5}) and Theorem \ref{theorem4.6}, we now propose the following problem.

\begin{problem}\label{problem5.1}
(a) Suppose $q\neq 2,3$. For $k=q$, do $k\times k$ NSC quasi-unitary matrices exist over $\mathbb{F}_{q^{2}}$?

(b) Suppose $q\neq 2$. For $q+2\leq k\leq q^{2}$, do $k\times k$ NSC quasi-unitary matrices exist over $\mathbb{F}_{q^{2}}$?
\end{problem}

We also propose the following problem.

\begin{problem}\label{problem5.2}
(a) How to find the NSC matrix $A$ over $\mathbb{F}_{q}$ such that $AA^{T}$ is monomial?

(b) How to find the NSC matrix $A$ over $\mathbb{F}_{q^{2}}$ such that $AA^{\dag}$ is monomial?
\end{problem}

\begin{remark}\label{remark5.4}
(a) To the best of our knowledge, a general method for finding such matrices is still lacking at present.
In a sense, once Problem \ref{problem5.2} is solved, a wider variety of good quantum codes (including EAQECCs) with parameters better than many
existing ones will be yielded.

(b) It will be interesting and challenging to explore the enumeration problems on some special matrices over finite fields, such as NSC matrices,
NSC quasi-orthogonal matrices, NSC quasi-unitary matrices and those proposed in Problem \ref{problem5.2}.
\end{remark}

\small{
\bibliographystyle{plain}
\phantomsection
\addcontentsline{toc}{section}{References}
\bibliography{ref2022.10.30}
}

\end{document}